\documentclass[12pt]{emulateapj}

\usepackage{amsmath,amssymb,natbib,graphicx}

\shorttitle{A Weak Lensing Study of Halo Centering}
\shortauthors{George et al.}
 
\begin{document}
  

 \title{Galaxies in X-ray Groups. II. A Weak Lensing Study of Halo Centering}



\author{
Matthew R. George\altaffilmark{1,2},
Alexie Leauthaud\altaffilmark{3},
Kevin Bundy\altaffilmark{3},
Alexis Finoguenov\altaffilmark{4,5},
Chung-Pei Ma\altaffilmark{1}, 
Eli S. Rykoff\altaffilmark{2,6}, 
Jeremy L. Tinker\altaffilmark{7},
Risa H. Wechsler\altaffilmark{6,8},
Richard Massey\altaffilmark{9},
Simona Mei\altaffilmark{10,11}
}

\altaffiltext{1}{Department of Astronomy, University of California,
  Berkeley, CA 94720, USA}

\altaffiltext{2}{Lawrence Berkeley National Laboratory, 1 Cyclotron
  Road, Berkeley, CA 94720, USA}

\altaffiltext{3}{Kavli Institute for the Physics and Mathematics of
  the Universe, Todai Institutes for Advanced Study, the University of
  Tokyo, Kashiwa, Japan 277-8583 (Kavli IPMU, WPI)}

\altaffiltext{4}{Max-Planck-Institut f\"{u}r Extraterrestrische
  Physik, Giessenbachstra\ss{}e, 85748 Garching, Germany}

\altaffiltext{5}{University of Maryland Baltimore County, 1000 Hilltop
  Circle, Baltimore, MD 21250, USA}

\altaffiltext{6}{Kavli Institute for Particle Astrophysics and
  Cosmology, SLAC National Accelerator Laboratory, 2575 Sand Hill
  Rd., Menlo Park, CA 94025, USA}

\altaffiltext{7}{Center for Cosmology and Particle Physics, Department
of Physics, New York University, 4 Washington Place, New York, NY
10003, USA}

\altaffiltext{8}{Physics Department, Stanford University, Stanford, CA
94305, USA}

\altaffiltext{9}{Department of Physics, University of Durham, South
  Road, Durham DH1 3LE, United Kingdom}

\altaffiltext{10}{University of Paris Denis Diderot, 75205 Paris Cedex 13, France}

\altaffiltext{11}{GEPI, Observatoire de Paris, Section de Meudon, 92195 Meudon Cedex, France}

\email{mgeorge@astro.berkeley.edu}


  
\begin{abstract}
  Locating the centers of dark matter halos is critical for
  understanding the mass profiles of halos as well as the formation and
  evolution of the massive galaxies that they host. The task is observationally
  challenging because we cannot observe halos directly, and tracers
  such as bright galaxies or X-ray emission from hot plasma are
  imperfect. In this paper we quantify the consequences of miscentering on
  the weak lensing signal from a sample of 129 X-ray selected galaxy
  groups in the COSMOS field with redshifts $0<z<1$ and halo masses in
  the range $10^{13}-10^{14}~{\rm M_{\odot}}$. By measuring the
  stacked lensing signal around eight different candidate centers
  (such as the brightest member galaxy, the mean position of all
  member galaxies, or the X-ray centroid), we determine which
  candidates best trace the center of mass in halos. In this sample of
  groups, we find that massive galaxies near the X-ray centroids
  trace the center of mass to $\lesssim 75$~{\rm kpc}, while the X-ray 
  position and centroids based on the mean position of member galaxies
  have larger offsets primarily due to the statistical uncertainties in their
  positions (typically $\sim50-150~{\rm kpc}$). Approximately $30\%$ of
  groups in our sample have ambiguous centers with multiple bright or
  massive galaxies, and some of these groups show disturbed mass
  profiles that are not well fit by standard models, suggesting that
  they are merging systems. We find that halo mass estimates from stacked
  weak lensing can be biased low by $5-30\%$ if inaccurate centers are
  used and the issue of miscentering is not addressed. 
\end{abstract}
 



\section{Introduction}

Galaxy groups and clusters are important sites of galaxy evolution and
the abundance of these massive objects provides a sensitive probe of
the amplitude of matter fluctuations and other cosmological
parameters. Analyses of these structures require some knowledge of the
location of the centers of their gravitational potentials. Because the
total mass distribution is dominated by dark matter and is not
directly observable, halo centers are typically assumed to be traced
by a massive galaxy or the density peak of radiating hot
gas. Miscentering is a critical issue when estimating the masses 
of groups and clusters because it adds significant systematic
uncertainties \citep[e.g.,][]{Johnston2007a, Johnston2007b,
  Mandelbaum2010, Rozo2011}, and also degrades constraints on the
concentration of mass profiles \citep{Mandelbaum2008}. Velocity
offsets between observational tracers and halo centers impact studies
of satellite kinematics \citep{Skibba2011, Wojtak2011} and
redshift-space distortions \citep{Hikage2012}. On the other hand, offsets between
observational tracers and the true halo centers can provide information about
the dynamical state of these systems and about the properties of dark
matter \citep{Clowe2006, Massey2011}.

Finding halo centers is challenging for a number of reasons. Galaxy
formation models (as well as halo models for describing the
multiplicity of galaxies within halos) typically place the brightest
or most massive galaxy at the center of each halo. But the brightest
galaxy in a cluster is not always the central galaxy \citep[][and
references therein]{Skibba2011}. Groups and clusters form from mergers
of halos where the most massive halo becomes the host halo with its
central galaxy, and smaller halos become subhalos with satellite
galaxies. Several analyses of data from group
    catalogs and field surveys have found that there is some intrinsic scatter
    in stellar mass and luminosity at fixed halo mass \citep{Yang2009,
      More2009, Leauthaud2012, Reddick2012}, which implies that
    halos can end up with satellites that are intrinsically more
    massive or luminous than the central galaxy. Additionally, there are
uncertainties in measuring any observable quantity such as stellar
mass that can cause a satellite to be misidentified as the most
massive central galaxy, and structure projected along the line of
sight can confuse the identification of member galaxies. Another difficulty
is that merging systems are dynamically unrelaxed, which can produce offsets
between the central galaxy and the halo center or other tracers such
as the X-ray center.  The systematics introduced by picking centers
that do not coincide with the ``true'' center of mass are important
and need to be quantified.

Gravitational lensing is a powerful tool for finding the centers of mass of halos
since it is sensitive to the total mass distribution along a line of
sight. Mass maps can be constructed for individual systems with strong
lensing or for massive clusters with weak lensing
\citep[e.g.,][]{Smith2005, Oguri2010, Shan2010}. Such studies often 
find reasonable agreement between the positions of bright massive
galaxies, X-ray emission, and lensing mass peaks, with a handful of
interesting examples that illustrate how dark matter peaks can be
offset from hot gas in merging galaxy clusters
\citep[e.g.,][]{Clowe2004, Bradac2008}. 

In this paper, however, we are concerned with a large statistical
sample of groups with lower masses and higher redshifts, a regime of
interest for many current and future weak lensing surveys. The typical
signal-to-noise ratio for the weak lensing signal from individual groups is
low, so we cannot identify their halo centers individually. By stacking
the lensing signal from many groups, we determine the mean mass
profile around a given center. We repeat this process for
different candidate centers and compare the resulting profiles to find
the best tracer of the center of mass.
The center of a smooth halo can be identified as the position where
the lensing signal is maximized on small scales. Other components such
as a massive galaxy or subhalo that is offset from the halo center
could produce an additional peak in the lensing signal, so we must
account for that when modeling the signal.

We analyze a sample of 129 X-ray selected galaxy groups at redshifts
$0<z<1$ from the COSMOS field \citep{Scoville2007a}, described in
\citet{George2011}.  With 
COSMOS data, we have X-ray luminosities and centroids for each group,
with member galaxies identified using photometric redshifts derived
from over thirty ultraviolet, optical, and infrared bands, and a
subsample with spectroscopic redshifts. We use weak lensing
measurements from high resolution \textit{Hubble} imaging to study the
accuracy with which tracers such as bright galaxies and X-ray emission
identify the centers of dark matter halos.

This paper is the second in a series studying the galaxy content of
X-ray groups. \citet[][hereafter Paper~I]{George2011} presented a
catalog of group membership assignments from a Bayesian treatment of
photometric redshifts, along with extensive tests of the selection
algorithm using mock catalogs and subsamples with spectroscopic
redshifts. Initial analyses of group members were used in that paper
to demonstrate an environmental influence on galaxy colors out to
$z=1$. A previous weak lensing study of this group sample was used to
constrain the mean relation between X-ray luminosity and halo mass
\citep{Leauthaud2010}. 

In this paper we study the centers of groups in
detail to optimize observational choices for centering, to study the
impact of miscentering on measurements of halo properties, and to
explore the effects of merging and substructure on lensing
measurements. The outline of the paper is as follows. Section~\ref{s:data} describes the data used in our
analysis, including the X-ray group catalog, assignment of member galaxies,
and lensing shape measurements. We define eight candidate group
centers in Section~\ref{s:centers}, and describe our procedure for
testing different choices of centers in Section~\ref{s:lensing}. Section~\ref{s:results}
presents the results of our analysis, and Sections~\ref{s:discussion}
and~\ref{s:conclusion} provide discussion and conclusions of our work.

We adopt a WMAP5 $\Lambda$CDM cosmology with $\Omega_{\rm m}=0.258$,
$\Omega_\Lambda=0.742$, $H_0=72$ $h_{72}$ km~s$^{-1}$~Mpc$^{-1}$
\citep{Dunkley2009} following the initial lensing analysis of these
groups by \citet{Leauthaud2010}. All distances are expressed in
physical units with $h=0.72$. X-ray luminosities are expressed in the 0.1-2.4 keV
band, rest-frame. All magnitudes are given on the AB system. To
approximate the virial radii of halos, we use $R_{\rm 200c}$ which is
the radius within which the mean mass density equals 200 times the
critical density of the Universe at the halo redshift, $\rho_{\rm
  c}(z)$. The corresponding mass enclosed within this radius is
$M_{\rm 200c}=200\rho_{\rm c}(4\pi/3)R_{\rm 200c}^3$. We also assume
halos follow a Navarro-Frenk-White \citep[NFW, ][]{Navarro1996} density
profile, with a concentration parameter $c_{\rm 200c}$ and a scale radius
$R_{s}=R_{\rm 200c}/c_{\rm 200c}$. We use the term ``group'' to 
denote a set of galaxies occupying a common halo, and the halo masses
of these groups is in the range $10^{13}-10^{14}~{\rm M_{\odot}}$ as
estimated with weak lensing \citep{Leauthaud2010}. We will generally
refer to more massive structures as clusters following convention, but
make no other physical distinction between groups and clusters.


\section{Data}
\label{s:data}

To study how the constituents of galaxy groups trace the centers of
mass of 
dark matter halos, we use an X-ray selected sample of galaxy groups
from the COSMOS field \citep{Scoville2007a}. We refer the reader to
Paper~I for details of the data and methods used to construct the
group catalog as well as tests of its properties with simulations and
spectroscopic data. In this section, we briefly describe aspects of the
catalog that are relevant for centering including the assignment of
member galaxies to groups. We also introduce the galaxy shear catalog
used in our weak lensing analysis.

\subsection{X-ray Group Catalog}
\label{s:xray}

Our sample of galaxy groups has been selected from an X-ray mosaic
combining images from the {\sl XMM-Newton} \citep{Hasinger2007} and
{\sl Chandra} \citep{Elvis2009} observatories following the procedure
of \citet{Finoguenov2009, Finoguenov2010}. A wavelet filtering of the
X-ray mosaic is used to distinguish extended structures on scales of
$32\arcsec$ and $64\arcsec$ from contaminants on smaller scales like
active galactic nuclei (AGN). Once extended X-ray sources are
detected, a red sequence finder is employed on galaxies with a
projected distance less than $0.5$~Mpc from the centers to identify an
optical counterpart and determine the redshift of the group, which is
then refined with spectroscopic redshifts when available.

A quality flag (hereafter \textsc{xflag}) has been assigned to each group
based on the reliability of the optical counterpart identification. We
study groups with \textsc{xflag}=1 or 2, indicating a confident spectroscopic
association, while higher values indicate uncertain counterparts which
could be due to projection effects or photometry contaminated by
bright foreground stars. Sources with \textsc{xflag}=1 also have clear
X-ray centroids, with an uncertainty in each position coordinate,
$\sigma_{\rm X}$, equal to the wavelet scale of $32\arcsec$ divided by
the signal-to-noise of the flux measurement, while sources with
\textsc{xflag}=2 have less certain X-ray centroids for which we have
$\sigma_{\rm X} = 32\arcsec$ set by the wavelet scale of the flux
measurement. 

To ensure a clean sample of groups with robust membership assignment,
we employ the additional quality cuts suggested in Paper~I, excluding
groups that are near field edges or have significantly masked areas,
potentially merging groups identified as distinct X-ray sources but
with significantly overlapping volumes, and poor groups with fewer
than four members identified. After these quality cuts, we have $129$
groups in our sample ranging from redshift $0<z<1$.

\subsection{Galaxy Membership}
\label{s:membership}

To determine how well galaxies trace the matter distribution in
groups, we must first identify the galaxies that reside in them. The
COSMOS field has extensive imaging in over thirty ultraviolet,
optical, and infrared bands \citep{Capak2007}, enabling the
determination of stellar masses (see Paper~I for details) and precise
photometric redshifts \citep[][and Paper~I for further
tests]{Ilbert2009}. In Paper~I, we presented a catalog of member
galaxies for these X-ray groups, selected according to their
photometric redshifts and proximity to X-ray centroids. Briefly, a
Bayesian membership probability, $P_{\rm mem}$, is assigned to each
galaxy by comparing the photometric redshift probability distribution
function to the expected redshift distribution of group and field
galaxies near each group. From the list of members with $P_{\rm mem}=1-P_{\rm
  field}>0.5$, the galaxy with the highest stellar mass within an NFW
scale radius of the X-ray centroid (including the positional
uncertainty, $\sigma_{\rm X}$) is selected as the group center. We
call this object the MMGG$_{\rm scale}$, for ``most massive group galaxy
within a scale radius''. A final membership probability is assigned by
repeating the selection process within a new cylinder re-centered on
this galaxy.

We have extensively tested this selection algorithm using mock
catalogs and with subsamples of galaxies for which we have
spectroscopic redshifts, and found it to be both pure and complete
near group centers; within $0.5 R_{\rm 200c}$ and down to our limiting
selection magnitude (F814W=24.2), $84\%$ of galaxies selected as group
members truly belong in groups, and $92\%$ of true group members are
selected as such. In this paper we use the member catalog derived from
photometric redshifts, which has an average of $26$ members per
group. From that list, there are an average of $6$ members per group
with spectroscopic redshifts for calibration and determining group
redshifts.

\subsection{Weak Lensing Data}

The galaxy shape measurements used for our weak lensing analysis are
described in \citet{Leauthaud2007}. These are derived from
high-resolution imaging over $1.64$~degrees$^2$ of the COSMOS field
with the Advanced Camera for Surveys (ACS) on the \textit{Hubble Space
  Telescope} \citep[\textit{HST};][]{Koekemoer2007} to a limiting magnitude of
F814W=26.4.  Variations in the point-spread function (PSF) with
position and time are modeled following \citet{Rhodes2007}, and galaxy
shapes are derived using the RRG method \citep{Rhodes2000}. The
PSF-corrected shapes are converted to estimators of shear, $\gamma$,
using a shear susceptibility factor calculated from moments of the
global distribution of shapes and a calibration factor determined from
simulated images. Updates to the procedure and the shear catalog are
described in detail elsewhere, in \citet{Leauthaud2012}. These improvements include a more
detailed correction of charge transfer inefficiency from
\citet{Massey2010}, and an empirical derivation of the dispersion in
shear measurements in bins of magnitude and detection
significance. This estimate of the shear dispersion includes
contributions from intrinsic shape noise and shape measurement
uncertainties, and varies from $\sigma_{\gamma}\approx 0.25$ for
bright galaxies to $\sigma_{\gamma}\approx 0.40$ for faint objects.

The stacked weak lensing signal is derived from the average
tangential shear, $\gamma_t(R)$, of background source galaxies at a
projected distance $R$ from the center of each group. The shear is related to
the excess surface mass density $\Delta\Sigma(R)$ \citep{Miralda1991}
\begin{equation}
\Delta\Sigma(R) \equiv \overline{\Sigma}(<R) - \overline{\Sigma}(R) =
\gamma_t(R)\Sigma_{\rm crit},
\label{eq:ds_def}
\end{equation}
where $\overline{\Sigma}(<R)$ is the mean surface density within
radius $R$ and $\overline{\Sigma}(R)$ is the azimuthally averaged
surface density at $R$. The critical surface density $\Sigma_{\rm
  crit}$ is a function of the angular diameter distances between the
observer ($O$), lens ($L$), and source ($S$),
\begin{equation}
\Sigma_{\rm crit}=\frac{c^2}{4\pi G}\frac{D_{OS}}{D_{OL}D_{LS}},
\label{eq:sigma_crit}
\end{equation}
where $c$ is the speed of light and $G$ is the gravitational constant.

In order to compute $\Sigma_{\rm crit}$ from
Equation~\eqref{eq:sigma_crit}, we need to know the distances to
both the sources and lenses. The group catalog provides lens redshifts which
come primarily from spectroscopic data including zCOSMOS \citep[][and
in prep.]{Lilly2009}. For background sources, we use photometric
redshifts from \citet{Ilbert2009}. To avoid contamination due to
uncertainties in photometric redshifts, we use only sources with
$z_S-z_L>\rm{max}[0.1,\sigma_{z}]$ where $\sigma_{z}$ is the $68\%$
uncertainty in the source redshift. We also exclude sources with a
secondary peak in their redshift density function (i.e. \textsc{zp}$_2
\neq 0$ in the \citealt{Ilbert2009} catalog) which have a significant
fraction of catastrophic redshift errors. With these cuts, the source
catalog contains $210,015$ galaxies with well-measured shapes and
redshifts, providing a source density of $36$ galaxies per arcmin$^2$.

To obtain a significant measurement of $\Delta\Sigma$, we must combine
the signal from many lenses and background sources. The combined
measurement is a weighted sum over pairs of lenses $i$ and sources
$j$,
\begin{equation}
\Delta\Sigma=\frac{\sum_i\sum_j\,\mathcal{W}_{ij}\gamma_{t,ij}\Sigma_{{\rm
      crit},ij}}{\sum_i\sum_j\,\mathcal{W}_{ij}}
\label{eq:ds_weighted}
\end{equation}
where the weight $\mathcal{W}_{ij}=(\Sigma_{\rm crit}
\sigma_{\gamma,ij})^{-2}$ is the inverse variance of the measurement.
We measure $\Delta\Sigma$ in annular bins from $20$~{\rm kpc} to
$1$~{\rm Mpc}. Covariance between measurements becomes an issue on
larger scales where background sources can be paired with multiple
lenses, but this is not significant over the scales we
measure. Uncertainties in $\Delta\Sigma$ are determined from the
inverse square root of the sum of the weights of lens-source pairs.


\section{Defining Candidate Centers}
\label{s:centers}

The ``center'' of a galaxy group requires some definition. There is
ambiguity in centering even when considering simulated dark matter
halos; the mass centroid, most bound particle, and density peak can
all be different because of asphericity and substructure. The
choice of group and cluster centers in observational data sets is
further limited by the available measurements. Here we review a
variety of definitions of group centers and their relative advantages.
Our aim is to use weak lensing to determine which candidates most
accurately trace (on average) the centers of mass of dark matter halos. We
will consider a variety of candidate centers and begin by studying the
level of agreement between them. Our choice of candidate centers is
meant to explore the range of options available for multi-wavelength
data sets while using a simple set of rules for identification;
however it is not an exhaustive list of possible centers.
 
It is useful to separate these definitions into two broad
categories. We call the first set ``galaxy candidates'' since they are
centered on a single galaxy, and the second set ``centroid
candidates'' which are defined for a spatially extended quantity like
the galaxy density field or X-ray emission and are in general not centered on an
individual galaxy. Some centering algorithms take a hybrid
approach, using the proximity of neighboring members to
ultimately select a luminous galaxy \citep[e.g.,][]{Robotham2011}, but we do
not test those methods here.

When identifying centers based on the galaxy content of groups, we
select from galaxies with membership probability $P_{\rm mem}>0.5$, as
described in Section~\ref{s:membership} and Paper~I. Though the list of
members is defined around one of the candidate centers (the MMGG$_{\rm
  scale}$), the radius ($R_{\rm 200c}$) used to select members is
large enough that the initial choice of center should not impact our
results. Each of the centers based on galaxy fluxes (e.g., brightest
group galaxy) use the observed magnitude in the F814W band, taken with
the ACS on \textit{HST}, with no corrections for dust or evolution. Since
these measurements do not account for the change in rest-frame
wavelength probed, they will be more sensitive to recent star
formation at higher redshifts. Centers based on galaxy masses use the
full measured spectral energy distribution (SED) so these effects are diminished.

\subsection{Galaxy Candidate Centers}
\label{s:gal_cand}

Many clusters have a central dominant galaxy with an extended stellar
envelope, often located near the density peak of hot intracluster gas
as seen in X-rays and the peak of the matter density probed by lensing
or kinematics \citep[e.g.,][and references therein]{Lin2004}. This
motivates the choice of a single galaxy to trace the centers of groups
and clusters. The general picture is further supported by numerical
simulations of dark matter halos and subhalos, and has been
encapsulated in the halo model which successfully describes many
aspects of large-scale structure including measurements of galaxy
clustering and lensing \citep[e.g.,][]{Cooray2002, Zehavi2005,
  Mandelbaum2006a, Leauthaud2012}.

Thus a popular choice when defining cluster centers in optical catalogs is
the Brightest Cluster Galaxy (BCG; or BGG in groups), since the
selection is relatively simple 
\citep[e.g.,][]{Koester2007a, Hao2010}. But the choice of filter and
aperture used for the flux measurement has an impact on which galaxies
are selected; differences in redshift, star formation history, and
dust content affect the flux observed in a given band, so a single
flux measurement cannot reflect the complicated physical processes
occurring in group centers. Color cuts can be used to isolate a few of
these effects \citep[e.g.,][]{Gladders2000}, though they often come
with assumptions about the properties of central galaxies. For example
some group catalogs use the brightest red sequence galaxy to identify
group centers, avoiding galaxies that are bright due to recent star
formation in favor of massive galaxies with old stellar
populations. 

Stellar masses are a promising alternative to observed or
rest-frame luminosities since they correlate more directly with the
masses of halos in which galaxies reside. However, stellar mass
estimates require more detailed measurements of the SEDs of galaxies
and are fraught with larger uncertainties than simple fluxes or luminosities.

In this paper we consider four galaxy candidates, selected based on
flux or stellar mass and distance to the X-ray position:

\begin{itemize}
\item MMGG$_{\rm scale}$: the galaxy within $R_{s}+\sigma_{\rm X}$ of
  the X-ray centroid having the greatest stellar mass.
\item MMGG$_{\rm R200}$: the galaxy having the greatest stellar mass
  of all group members within $R_{\rm 200c}$.
\item BGG$_{\rm scale}$: the brightest galaxy within $R_{s}+\sigma_{\rm X}$ of
  the X-ray centroid.
\item BGG$_{\rm R200}$: the brightest galaxy of all group members
  within $R_{\rm 200c}$.
\end{itemize}

The X-ray centroid (with uncertainty $\sigma_{\rm X}$) is used as the starting point for selecting these
galaxies because it should roughly trace the halo center and we do not
have lensing centers for individual groups. Note that there is not
necessarily a galaxy within the NFW scale radius $R_s$ of the X-ray
centroid, so MMGG$_{\rm scale}$ and BGG$_{\rm scale}$ are not
necessarily defined for all galaxy groups. However, in the case of our
clean sample, each group has at least one member within this radius so
we do not have to deal with undefined centers. Uncertainties in the
galaxy positions are much smaller than the sizes of the galaxies, and
are therefore negligible compared to the offsets from halo centers
that we are capable of measuring with weak lensing.

\subsection{Centroid Candidate Centers}

The central galaxy is not always observationally obvious, and
selection of an incorrect galaxy can produce statistically undesirable
results when studying a sample of groups or clusters. This problem has
motivated the use of centroids based on the positions of some or all
group members, which can be weighted by their properties such as flux
or stellar mass, with the hope that a robust statistic can be less
prone to large errors than the choice of a single galaxy
\citep[e.g.,][]{White1999, Carlberg2001, Berlind2006, Jee2011}.

Additionally, other probes of groups and clusters such as X-ray and SZ
\citep{Sunyaev1972} 
observations of hot gas, and gravitational lensing can be used to find
halo centers. Deep pointed observations can map the gas distribution
in great detail for bright systems that are nearby or massive, but
centering uncertainties can be significant for fainter systems (see
Section~\ref{s:xray}). Similarly, only very massive systems produce a
large enough lensing signal to study their spatial mass distribution
individually, and lower mass systems (like the groups studied here)
require stacking, such that centroids cannot be determined for each
individual group from lensing alone.

Here we consider four centroid candidates:
\begin{itemize}
\item CN: the centroid of member galaxies.
\item CM: the centroid of member galaxies weighted by stellar mass.
\item CF: the centroid of member galaxies weighted by flux
\item X-ray: the X-ray centroid.
\end{itemize}

Uncertainties on the X-ray positions were discussed in Section~\ref{s:xray}
and have a mean value of $22\arcsec$ or $134~\rm{kpc}$. For the other
centroid candidates (CN, CM, CF), the coordinates are computed using a
weighted mean,
\begin{equation}
\mathbf{x_{\rm cen}} = \frac{\sum\limits_{i=1}^{N} w_i \mathbf{x_i}}{\sum\limits_{i=1}^{N} w_i},
\end{equation}
where $\mathbf{x_i}$ is the pair of coordinates (R.A., Dec.)$_i$ for
each galaxy $i$, $N$ is the number of group members, and $w_i$ is the
appropriate galaxy weight for each center definition; $w_i=1,
M_{\star,i}, f_i$ for CN, CM, CF, respectively, where $M_{\star,i}$ is
the stellar mass and $f_i\propto10^{-0.4m_i}$ with $f_i$ and $m_i$ the
flux and apparent F814W magnitude for each galaxy. We estimate the
errors for these weighted means using bootstrap resampling from the
list of member galaxies.
With an average of 26 member galaxies per group, the mean statistical
uncertainties on the projected galaxy centroid positions are $45, 52,$ and
$50~\rm{kpc}$ for candidates CN, CM, and CF, respectively, where we
have taken the geometric mean of the uncertainties in two dimensions
and converted the angular uncertainty to a projected physical distance
at the redshift of each group. Groups with a higher projected density
of member galaxies tend to have smaller centroid uncertainties, but
improvements appear to be limited by contamination in the outskirts
from correlated structure (see Paper I for tests of purity and
completeness). We have tested different centroiding schemes including
iterating until the centroid and member list converge or restricting
to red galaxies, but achieved qualitatively similar results as with
the centroids presented.

\subsection{Offsets Between Candidate Centers}

\begin{figure*}[htb]
\plotone{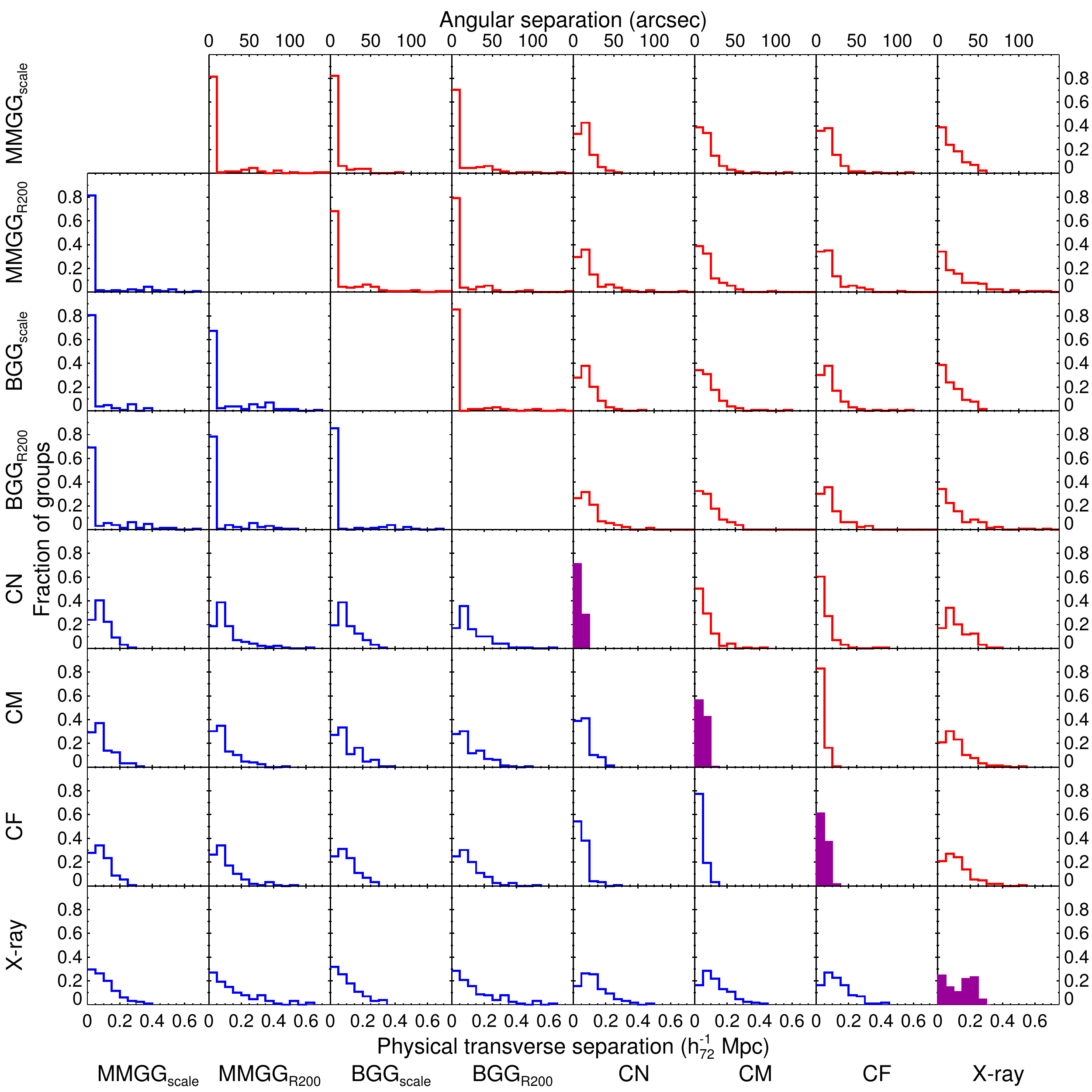}
\caption{Distribution of projected offsets between pairs of candidate
  centers in our group sample, measured in arcseconds (upper right;
  red) and ${\rm Mpc} $(lower left; blue). The angular and physical
  offset distributions are not identical because the groups span a
  range of redshifts. The filled purple histograms on the diagonal
  panels show the distribution of statistical uncertainties for each
  centroid position, described in Section~\ref{s:xray} for the X-ray
  centroid and Section~\ref{s:gal_cand} for the others. The y-axis
  gives the fraction of groups in each bin; 
  bin sizes are $50$~\rm{Mpc} (bottom left and diagonal) and
  $10$\arcsec (upper right).}
\label{fig:offsets}
\end{figure*}

Our first test of these various centers is to see how well they agree with one
another. Figure~\ref{fig:offsets} shows the distribution over all
groups of the angular and physical distance offsets between pairs of
candidate centers, along with the distribution of uncertainties in
centroid positions. Immediately we see that candidate centers do not 
always agree. For example, in $22\%$ of groups the brightest galaxy
within $R_{\rm 200c}$ is not the most massive galaxy and the
candidates are separated by up to several hundred~{\rm kpc}. The
agreement level among pairs of galaxy candidates is typically
$70-80\%$ with a long tail in the distribution extending out roughly
to the virial radius for these groups. The galaxy candidates are
typically offset from the centroid candidates by $50-100$~{\rm kpc},
again with tails of a few hundred~{\rm kpc}, and the centroid
candidates are in slightly better agreement among themselves. The
offsets between the X-ray centroid and other candidate centers are
generally consistent with the statistical uncertainties on the X-ray
centroid. When comparing galaxy centroids (CN, CM, and CF) to other
candidate centers, the typical offsets are roughly consistent with the
mean uncertainty on the centroid position, but there are long tails
in the offset distribution that exceed typical uncertainties.

These results are generally consistent with offsets found in other
groups and cluster samples, though direct comparison is difficult
given the variety of methods and data used for identifying objects and
their centers. For $\sim30\%$ of optically-selected groups in a
similar mass range as our sample, \citet{Skibba2011} found that the
brightest galaxy was not the central one, based on the relative positions
and velocities of other member galaxies. This is comparable to the
level of disagreement we find between our galaxy candidates, for which
choosing a central galaxy is ambiguous. Comparing the positions of
BCGs to X-ray centroids in 42 optically-selected clusters,
\citet{Sheldon2001} noted a mean offset of $85~{\rm kpc}$. With an
expanded sample of 94 clusters with matching X-ray detections,
\citet{Koester2007b} found a very similar median BCG-X-ray offset of
$81~{\rm kpc}$, and noted $\sim20\%$ of systems with offsets of
several hundred kpc which were mostly due to confusion in identifying
the X-ray position or BCG. In a study of 65 massive clusters with
higher quality X-ray data, \citet{Sanderson2009} found BCG-X-ray
offsets of typically less than a few tens of kpc, with a few
outliers that were merging systems. Our X-ray offsets are somewhat
larger due to the statistical uncertainties in the centroid positions,
with a mean (median) offset of $104$~{\rm kpc} ($78$~{\rm kpc})
between the MMGG$_{\rm scale}$ and X-ray centroid.

\section{Weak Lensing Methodology}
\label{s:lensing}

\subsection{The Approach}

\begin{figure*}[htb]
\epsscale{0.9}
\plotone{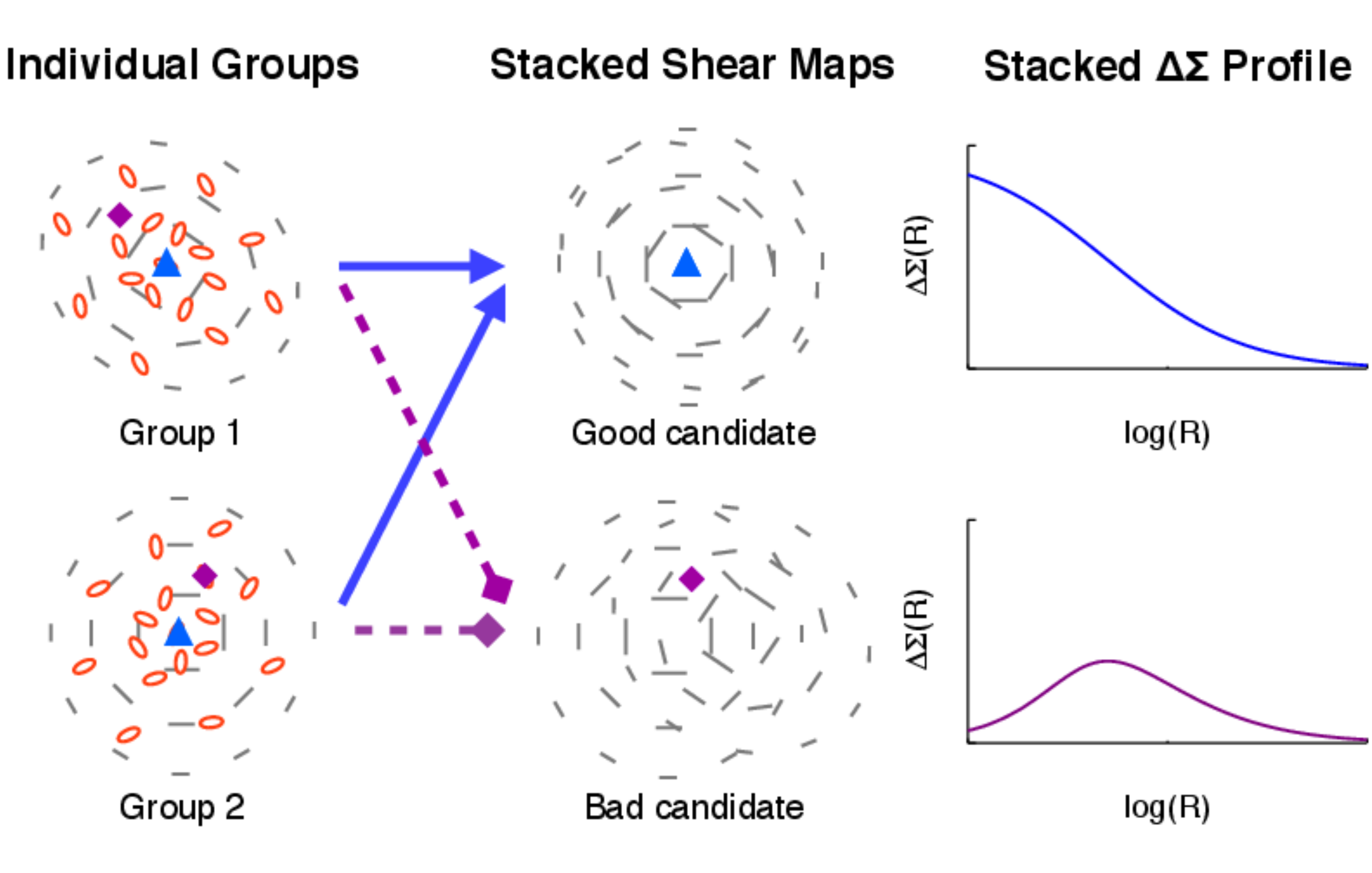}
\caption{Schematic illustration of stacked lensing around different
  candidate centers. Candidate centers are defined in each group
  (left), then shear maps are stacked around each position (middle), and
azimuthally averaged to compute $\Delta\Sigma$ profiles (right).}
\label{fig:cartoon}
\end{figure*}

Our stacked weak lensing approach to test candidate centers is
sketched in Figure~\ref{fig:cartoon}. The left column shows two
separate galaxy groups (red ellipses) and their corresponding shear
maps measured from the shapes of background galaxies (gray
sticks). For each group, two candidate centers are defined (blue
triangle and purple diamond), and the shear maps are stacked around
these positions in the middle column. The rightmost panels show the
resulting lensing signal $\Delta\Sigma$ as measured radially from the
candidate center. We emphasize that the lensing signal for individual
groups studied in this paper is noisy \citep[signal-to-noise $\sim1$;
see Figure 1 of ][]{Leauthaud2010}, so we cannot directly identify the
centers of weak lensing maps for individual groups and must stack many
groups; in this sense Figure~\ref{fig:cartoon} is exaggerated.

Qualitatively, the amplitude of the lensing signal is maximized when
the lens position used for stacking coincides with the true center of
mass in each system, and the signal is suppressed when the nominal
position deviates from the true center of mass. The two curves agree
at radii much larger than the typical centering offset.  More
formally, the lensing signal around miscentered halos was first
studied in the context of satellite galaxies \citep{Natarajan1997,
  Hudson1998, Guzik2002, Yang2003, Yang2006}, and later applied to the
problem of uncertain group centers \citep{Johnston2007a,
  Johnston2007b}.

Observationally, our aim is to find the candidate center that
maximizes the lensing signal on small scales, indicating that it best
traces the center of mass. Furthermore, we would like to model the
signal to infer the underlying mass profile and the typical offsets
between our tracers and the true center of mass in halos. Interpreting
the signal is complicated by a number of effects including the shapes
of halo profiles and the properties of galaxies and subhalos which
will be discussed further in Section~\ref{s:discussion}.

\subsection{Models}

To interpret the mean lensing signal $\Delta\Sigma$ defined in
Equations~\eqref{eq:ds_def} and~\eqref{eq:ds_weighted}, we construct a
model for the surface mass density $\Sigma(R)$ of a typical
lens. Contributions to $\Sigma(R)$ in our group sample come primarily from
the dark matter halo and the central galaxy (if the center is defined
to be at the position of a galaxy). The parameters of the model are introduced in
  Table~\ref{tab:modelPars}. We include three variants of the model: a
\textit{centered} version where the halo center is fixed at the position of the
candidate center, an \textit{offset} model which allows a distribution
of offsets between the candidate and true halo center, and the
\textit{full} model which adds freedom to the halo profile and allows
for excess mass in the form of a subhalo around the candidate center.

\begin{deluxetable*}{llcccccc}
\tablecaption{Model Parameters for $\Delta\Sigma(R)$}
\tablehead{\colhead{Parameter} & \colhead{Description} & \colhead{Centered}
  & \colhead{Offset} & \colhead{Full} & \colhead{Prior mean} &
  \colhead{Prior $\sigma$} & \colhead{Restrictions}
}
\startdata
$\log(M_{\rm 200c}/M_{\odot})$ & halo mass & free & free & free & $13.5$ & $0.8$ & - \\
$c_{\rm 200c}$ & halo concentration & tied & tied & free & $4.0$ & $3.0$ & $1<c_{\rm 200c}<10$ \\
$\sigma_{\rm off}~(\rm{kpc})$ & offset distance scale & fixed & free & free & $0$ & $200$ & $\sigma_{\rm off}>0$ \\
$\log(M_{\rm gal}/M_{\odot})$ & stellar mass  & fixed & fixed & fixed & $\langle\log(M_{\rm \star, gal}/M_{\odot})\rangle$ & - & - \\
$\log(M_{\rm sub}/M_{\odot})$ & subhalo mass & omitted & omitted & free & $\langle\log(M_{\rm \star, gal}/M_{\odot})\rangle$ & $1.0$ & $\log(M_{\rm sub}/M_{\odot})>10$
\enddata
\tablecomments{For each model, $M_{\rm \star, gal}$ is
    fixed to the mean photometrically-estimated stellar mass of the
    central galaxy (for galaxy candidates) and zero otherwise. For the
    centered and offset models, halo concentration is set by the
    relation of \citet{Zhao2009}. \label{tab:modelPars}}
\end{deluxetable*}

 We model the average mass density in halos with a spherical NFW profile,
for which the projected surface density $\Sigma_{\rm halo}(R)=\Sigma_{\rm NFW}(R)$ is given in
e.g. \citet{Wright2000}, with halo mass and concentration as two free
parameters. For the centered and offset models we will assume a
mass-concentration relation from \citet{Zhao2009}, leaving 
mass as a single free parameter for the halo
component, while both mass and
  concentration are free parameters in the full model.

When the surface density of a spherically symmetric halo is measured
around the correct center of mass, we have
$\overline{\Sigma}(R)=\Sigma_{\rm halo}(R)$. If there is an offset
$R_{\rm off}$ in the lens plane between the true center and the
position used for measurement, the surface density measured at the
coordinates ($R, \theta$) relative to the offset position is
\citep[][Appendix B]{Yang2003}
\begin{equation}
\Sigma_{\rm halo}^{\rm off}(R,\theta|R_{\rm off})=\Sigma_{\rm halo}\left(\sqrt{R^2+R_{\rm off}^2-2RR_{\rm off}\cos\theta}\right).
\end{equation}
The azimuthally-averaged surface density around the offset position is
\begin{equation}
\overline{\Sigma}_{\rm halo}^{\rm off}(R|R_{\rm off})=\frac{1}{2\pi} \int_0^{2\pi}\mathrm{d}\theta\,
\Sigma_{\rm halo}^{\rm off}(R,\theta|R_{\rm off}).
\label{eq:Sigma_R_Roff}
\end{equation}
For an ensemble of halos with a distribution of offsets 
$P(R_{\rm off})$, \citet{Johnston2007a, Johnston2007b} generalized
Equation~\eqref{eq:Sigma_R_Roff} to give the mean azimuthally-averaged
surface mass profile stacked around the offset positions
\begin{equation}
\overline{\Sigma}_{\rm halo}^{\rm off}(R|P(R_{\rm off}))=\int_0^{\infty} 
P(R_{\rm off})\,\overline{\Sigma}_{\rm halo}^{\rm off}(R|R_{\rm
  off})\,\mathrm{d}R_{\rm off}.
\end{equation}
The mean surface density inside a radius $R$ is
\begin{align}
\overline{\Sigma}_{\rm halo}^{\rm off}(<R|P(R_{\rm off})) &=
\frac{1}{\pi R^2}
\int\limits_0^R\int\limits_0^{2\pi}\int\limits_0^{\infty} P(R_{\rm off})\,\Sigma_{\rm halo}^{\rm off}(R',\theta|R_{\rm off})\nonumber\\
& \qquad \qquad \qquad \qquad \quad \times \,R'\mathrm{d}R'\mathrm{d}\theta\mathrm{d}R_{\rm off} \nonumber\\
&= \frac{2}{R^2} \int_0^R \overline{\Sigma}_{\rm halo}^{\rm off}(R'|P(R_{\rm off}))\,R'\mathrm{d}R'.&
\end{align}

To model the lensing signal from a large sample of galaxy clusters
centered around BCGs, \citet{Johnston2007b} used a distribution of
offsets $P(R_{\rm off})$ estimated from mock catalogs. In their model
a fraction of BCGs correctly identified the centers of halos
($R_{\rm off}=0$), while the remaining clusters had a distribution of
offsets given by
\begin{equation}
P(R_{\rm off})=\frac{R_{\rm off}}{\sigma_{\rm off}}
\exp\left(-\frac{R_{\rm off}^2}{2 \sigma_{\rm off}^2}\right).
\label{eq:rayleigh}
\end{equation}
This model, called a two-dimensional Gaussian or a Rayleigh
distribution, was chosen based on mock catalogs to which their
cluster-finding algorithm had been applied. The mocks suggested
that the fraction of correctly-centered clusters depended on
richness, with higher richness clusters more likely to be centered
correctly. In other clusters, the central galaxy was not correctly
identified as the BCG, and the distribution of offsets between the BCG
and true halo center could be described by
Equation~\eqref{eq:rayleigh} with the parameter $\sigma_{\rm
  off}=420~h^{-1}~\rm{kpc}$ describing the typical offset scale,
independent of cluster richness.

We can think of the offset more generally in three dimensions, where
we assume the offset in each dimension is normally distributed with
mean zero. The observed offset in the line-of-sight dimension might
not have the same variance as the dimensions in the lens plane because
of redshift-space distortions, but as long as the distribution in
three dimensions is joint-normal and the variance in the two
dimensions of the lens plane is equal, the projected offset distribution
will take the form of Equation~\eqref{eq:rayleigh} after marginalizing
over the line of sight.

The miscentering component employed by \citet{Johnston2007b} added
two free parameters to their model: the fraction of miscentered
groups, and the scale length of the offset distribution. These
additional parameters could not be well-constrained by the data, and
had to be constrained using strong priors from mock catalogs.

Our approach is to use an offset model with a single free parameter to
constrain the scale of the miscentering distribution empirically.  We
do not include separate components for centered and miscentered groups
as \citet{Johnston2007b} did, as the single parameter model generally
provides sufficient fits to the current data. Also, correctly chosen central
galaxies may still be offset from their halo centers, an effect that
was not included in the mocks of \citet{Johnston2007b} but has been
tentatively seen in lensing maps of individual clusters
\citep{Oguri2010}. 

The second component of the model is the surface density contributed by
the central galaxy, $\Sigma_{\rm gal}(R)$. The shape of this mass
profile is uncertain, but its contribution is subdominant even
at the smallest radii that we probe ($R\sim50~{\rm
  kpc}$). For the centered and offset models, we simply
model the galaxy component as a point source, $\Sigma_{\rm gal}(R)=M_{\rm gal}/(\pi
R^2)$, with $M_{\rm gal}$ fixed to the average stellar mass $M_{\rm
  \star,gal}$ of the central galaxies in the ensemble as estimated
from their SEDs. The centroid candidates (CN, CM, CF, and
X-ray) typically do not have a galaxy very near the center, so we do
not include a contribution from $\Sigma_{\rm gal}$ when modeling the
signal from these candidates.

We can write the centered and offset models as the sum of the halo and
galaxy components:
\begin{eqnarray}
\Delta\Sigma^{\rm cen}(R) &=& \Delta\Sigma_{\rm NFW}(R) + \Delta\Sigma_{\rm gal}(R) \\
\Delta\Sigma^{\rm off}(R) &=& \Delta\Sigma_{\rm NFW}^{\rm off}(R|P(R_{\rm off})) + \Delta\Sigma_{\rm gal}(R).
\end{eqnarray}
Note that $\Delta\Sigma^{\rm cen}(R)=\Delta\Sigma^{\rm off}(R|R_{\rm
  off}=0)$.  We later test a scenario with the full model in which the
candidate center has an additional mass component in the form of a
dark matter subhalo that is more extended than the stellar profile
which will be described further in Section~\ref{s:model_unc}.

Though the measured signal is an
ensemble average coming from halos with a range of masses and mass
profiles, our model consists of a single profile for simplicity and
because the range of halo masses inferred from group X-ray
luminosities is relatively small. We restrict our analysis of the
lensing signal to $R<1~{\rm Mpc}$, where the effects of halo
truncation and correlated structure should be unimportant to the
lensing signal, and note that this adequately covers the range of
centering offsets shown in Figure~\ref{fig:offsets}.
At small scales, the assumption of weak shear becomes less
accurate. \citet{Mandelbaum2006b} derive a correction term to the
surface density contrast that depends on $\Sigma_{\rm crit}$ and
$\Sigma(R)$ for the sample. These correction factors have been
computed for this sample by \citet{Leauthaud2010} and shown to be
fairly small on the scales we probe (of order $10\%$ of the measured
signal in our innermost bin for a good center), so we do not include
them in our analysis.

To fit the models to the data, we attempt to find the
  parameters that minimize
\begin{equation}
\chi^2=\sum_i \frac{\left(\Delta\Sigma_{\rm data}(R_i)-\Delta\Sigma_{\rm model}(R_i)\right)^2}{\sigma_i^2}
\end{equation}
where $\sigma_i$ is the measurement uncertainty on
  $\Delta\Sigma_{\rm data}(R_i)$. In practice, we use a Markov Chain
  Monte Carlo (MCMC) approach to efficiently explore the multi-dimensional
  parameter space maximizing the logarithm of the likelihood,
  $L\propto\exp(-\chi^2/2)$. We employ Gaussian or lognormal priors for each
  parameter, with means and standard deviations given in
  Table~\ref{tab:modelPars}. The data are unable to constrain a lower
  limit to the subhalo mass or an upper limit to the concentration, so we restrict these
parameters to $\log(M_{\rm sub}/M_{\odot})>10$ and $c_{\rm
  200c}<10$ when they are free. We also require $c_{\rm 200c}>1$ and $\sigma_{\rm
  off}>0$.


\section{Results}
\label{s:results}

\subsection{Weak Lensing on the Full Sample}
\label{s:comparing_cen}

Given that Figure~\ref{fig:offsets} shows a wide range of
offsets between different choices of group centers, we proceed
to test how well the candidate centers trace the underlying matter
distribution. We begin by studying the full sample of groups in this
section, and in the next section we focus on the subset of groups with significant
offsets between candidates. 

\begin{figure*}[htb]
\epsscale{1.1}
\plotone{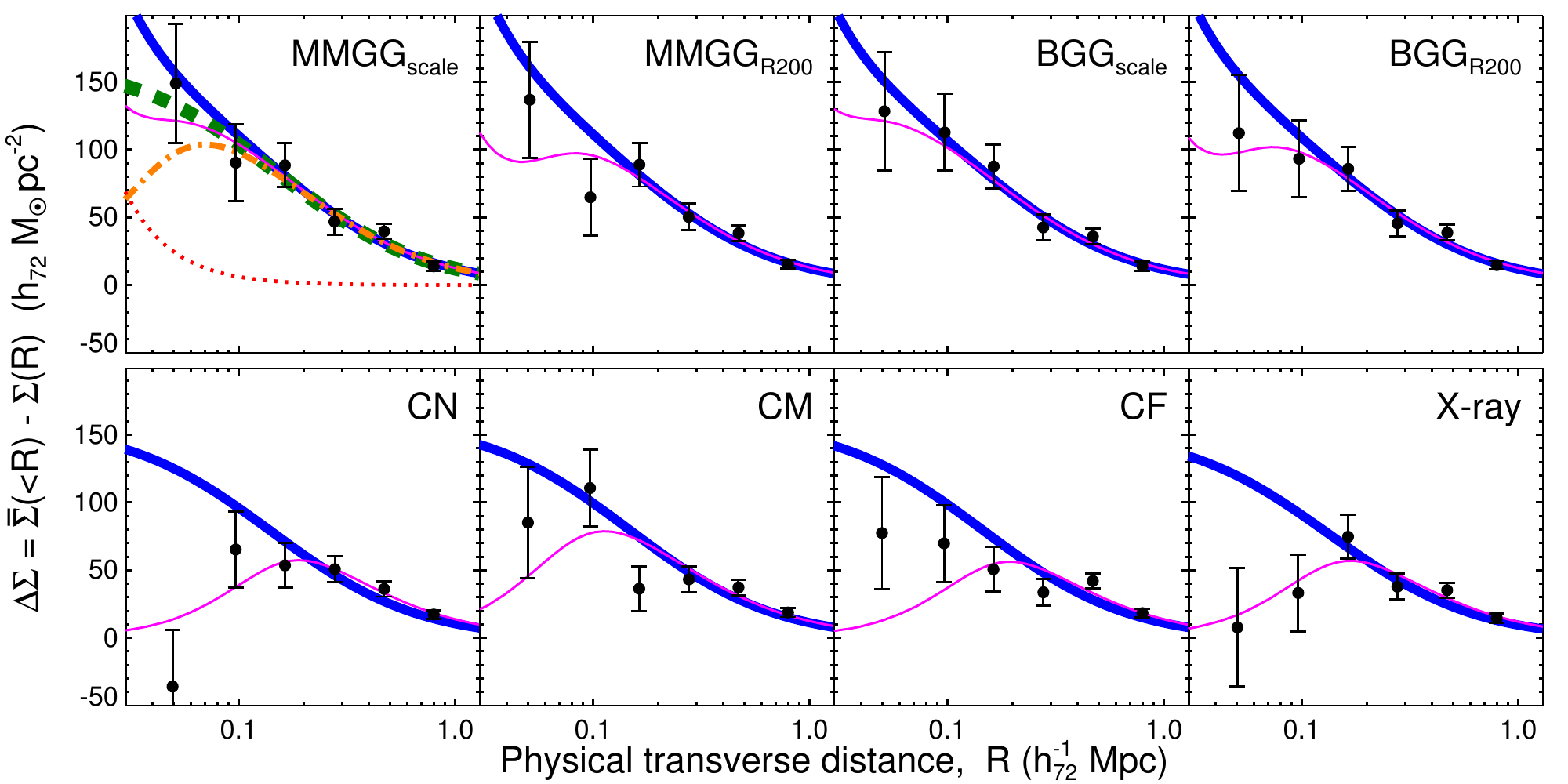}
\caption{Weak lensing signal stacked on the full sample of groups
  around different centers, along with centered (thick blue,
  $\Delta\Sigma^{\rm cen}$) and offset (thin magenta,
  $\Delta\Sigma^{\rm off}$) models. Halo and central components of
  these models are shown for MMGG$_{\rm scale}$ (green dashed for
  $\Delta\Sigma^{\rm cen}_{\rm NFW}$, orange dot-dashed for
  $\Delta\Sigma^{\rm off}_{\rm NFW}$, red dotted for
  $\Delta\Sigma_{\rm gal}$). The top row shows the signal around
  galaxy candidates, while the bottom row shows centroid
  candidates. The signal (black points) is measured in radial bins
  with the first spanning $20-70$~{\rm kpc} for sufficient
  signal-to-noise, then logarithmically spaced from $70$~{\rm kpc} to
  $1$~{\rm Mpc}.}
\label{fig:full_stacks}
\end{figure*}

For each center candidate, we measure the lensing signal
$\Delta\Sigma(R)$ from all groups in annular bins around the center.
The results are shown for the eight candidate centers in
Figure~\ref{fig:full_stacks}. Each panel represents a different
candidate center from Section~\ref{s:centers}, for which we plot the
measured $\Delta\Sigma(R)$ (black points) along with models
$\Delta\Sigma^{\rm cen}(R)$ (thick blue) and $\Delta\Sigma^{\rm
  off}(R)$ (thin magenta). For illustration, we show the halo and
galaxy components of the models for MMGG$_{\rm scale}$:
$\Delta\Sigma_{\rm NFW}(R)$ (green dashed), $\Delta\Sigma_{\rm
  NFW}^{\rm off}(R|R_{\rm off})$ (orange dot-dashed), and
$\Delta\Sigma_{\rm gal}(R)$ (red dotted). 

The mean and standard deviation ($1\sigma$) of the
  posterior distribution for each parameter, along with the mean
central galaxy masses and $\chi^2$ values for the fits, are shown in
Table~\ref{tab:fitPars}. The $\chi^2$ values have not
  been normalized by the number of degrees of freedom $\nu$. Each
  model is fit to $6$ data points with $1$ and $2$ free parameters for
  the centered and offset models, respectively, so $\nu=5$ for the
  centered model and $\nu=4$ for the offset model. When $\chi^2
  \approx \nu$, the model is consistent with the data given the
  uncertainties. A high value of $\chi^2$ indicates that the model
  does not fit the data; for example, the data are inconsistent with
  the model at a $95\%$ confidence level when $\chi^2 \geq
  11.07~(9.49)$ for $\nu= 5~(4)$.

\begin{deluxetable*}{lccccccccc}
\tablecaption{Parameters Constraints for $\Delta\Sigma(R)$}
\tablehead{ \colhead{} & \colhead{} & \multicolumn{3}{c}{Centered
    Model} & \colhead{} & \multicolumn{4}{c}{Offset Model} \\
  \cline{3-5} \cline{7-10} \\
  \colhead{Candidate} & \colhead{$\langle\log(M_{\rm \star, gal}/M_{\odot})\rangle$} &
  \colhead{$\log(M_{\rm 200c}/M_{\odot})$} & \colhead{$c_{\rm 200c}$}  & \colhead{$\chi^2$ $(\nu=5)$} & &
  \colhead{$\log(M_{\rm 200c}/M_{\odot})$} & \colhead{$c_{\rm 200c}$}  & \colhead{$\sigma_{\rm off}$ (kpc)} & \colhead{$\chi^2$ $(\nu=4)$}
}
\startdata
\sidehead{\textit{Galaxy candidates}}
  MMGG$_{\rm scale}$ & 11.3 & 13.43 $\pm$  0.07 &  4.0 &  4.2 & &  13.45 $\pm$  0.07 &  3.9 &  18.2 $\pm$  11.3 &  4.2 \\
    MMGG$_{\rm R200}$ & 11.4 & 13.43 $\pm$  0.06 &  4.0 &  5.3 & &  13.47 $\pm$  0.07 &  3.9 &  28.3 $\pm$  15.2 &  4.4 \\
    BGG$_{\rm scale}$ & 11.2 & 13.40 $\pm$  0.06 &  4.0 &  2.9 & &  13.43 $\pm$  0.07 &  4.0 &  16.8 $\pm$  10.0 &  2.8 \\
     BGG$_{\rm R200}$ & 11.3 & 13.42 $\pm$  0.07 &  4.0 &  4.4 & &  13.45 $\pm$  0.06 &  3.9 &  24.8 $\pm$  12.0 &  3.0 \\
\sidehead{\textit{Centroid candidates}}
             CN &    - & 13.35 $\pm$  0.07 &  4.0 & 21.0 & &  13.51 $\pm$  0.07 &  3.9 &  65.0 $\pm$  18.0 &  2.8 \\
             CM &    - & 13.39 $\pm$  0.07 &  4.0 & 10.5 & &  13.45 $\pm$  0.08 &  3.9 &  34.5 $\pm$  21.6 &  9.3 \\
             CF &    - & 13.38 $\pm$  0.07 &  4.0 & 14.0 & &  13.51 $\pm$  0.09 &  3.9 &  66.9 $\pm$  29.9 &  9.5 \\
          X-ray &    - & 13.30 $\pm$  0.08 &  4.1 & 15.2 & &  13.42 $\pm$  0.08 &  4.0 &  57.1 $\pm$  16.9 &  3.5
\enddata
\tablecomments{For both models, $M_{\rm \star, gal}$ is fixed to the
  stellar mass of the central galaxy (for galaxy candidates) and zero
  otherwise, and the halo concentration is fixed by the relation of
  \citet{Zhao2009}. \label{tab:fitPars}}
\end{deluxetable*}

There are clear differences among the lensing signals for the eight
candidate centers. The signal generally rises toward small scales for
the galaxy candidates, while there is a turnover for the centroid
candidates indicating that they are poorer at tracing the center of
mass in these groups. The stellar mass in the candidate central
galaxies produces some of the difference in the signal, but the
best-fit models show that the galaxy candidates have smaller offsets
from the halo centers when compared with the centroid candidates even
when accounting for the stellar mass. We will explore the possibility
of additional mass in the central galaxy candidates in
Section~\ref{s:model_unc}. It is important to note that the offsets
measured from the lensing signals for the X-ray position and other
centroid candidates are primarily due to the large statistical
uncertainties in the centroid positions, shown in
Figure~\ref{fig:offsets}. Figure~\ref{fig:full_stacks} shows the
impact of these centroid uncertainties on the lensing signal, but we
cannot infer from this data that there are significant intrinsic
offsets between the true centroid positions and the center of the dark
matter halo.

For each of the galaxy candidates, the $\chi^2$ value
  in Table~\ref{tab:fitPars} indicates that the centered model
  provides a good description of the data given the error bars. The
  extra degree of freedom in the offset model does not significantly
  improve the fit, and the constraints on $\sigma_{\rm off}$ are
  consistent with zero at the $2\sigma$ level. The best-fit values for
  $\sigma_{\rm off}$ and the quality of the fits are statistically
  consistent for each of the galaxy candidates, so we do not strongly
  favor one candidate over another, though there is marginal evidence
  that the candidates defined within an NFW scale radius of the X-ray
  position (MMGG$_{\rm scale}$ and BGG$_{\rm scale}$) trace the halo
  center slightly better than candidates defined over the larger area
  within $R_{\rm 200c}$.
 
On the other hand, the centered model generally does not provide good
fits to the data for the centroid candidates, and the offset models do
improve the fits. The best-fit offset parameter
  $\sigma_{\rm off}$ is significantly larger for the centroid candidates than for the galaxy
  candidates and deviates from zero by more than $2\sigma$ for three
  of the centroid candidates. The offset model fits
  two centroid candidates (CM and CF) only marginally well so these
would perhaps be better fit by a more complicated offset distribution.

The effect of miscentering on the halo mass constraints is comparable
to the statistical uncertainty for this sample. The best-fit halo
masses assuming a centered model for the centroid candidates tend to
be $\sim1\sigma$ lower than for galaxy candidates. Halo masses
increase slightly when offsets are allowed in the model, and
candidates with larger offsets are more affected. When miscentering is
not accounted for in the model, halo masses are underestimated by
$5-30\%$ compared to when we include the effect, depending on the
choice of center.

We have also tested the effect of fixing the concentration parameter
with our assumed mass-concentration relation, and freeing this
parameter does not qualitatively change the results. We do not obtain
a good constraint on the concentration, but the best-fit value for the
galaxy candidates is consistent with our assumed value. The centroid
candidates prefer somewhat lower concentrations, but the values are
still consistent within the large error bars.

\subsection{Groups with Discrepant Candidate Centers}
\label{s:discrepant}

\begin{figure*}[htb]
\plotone{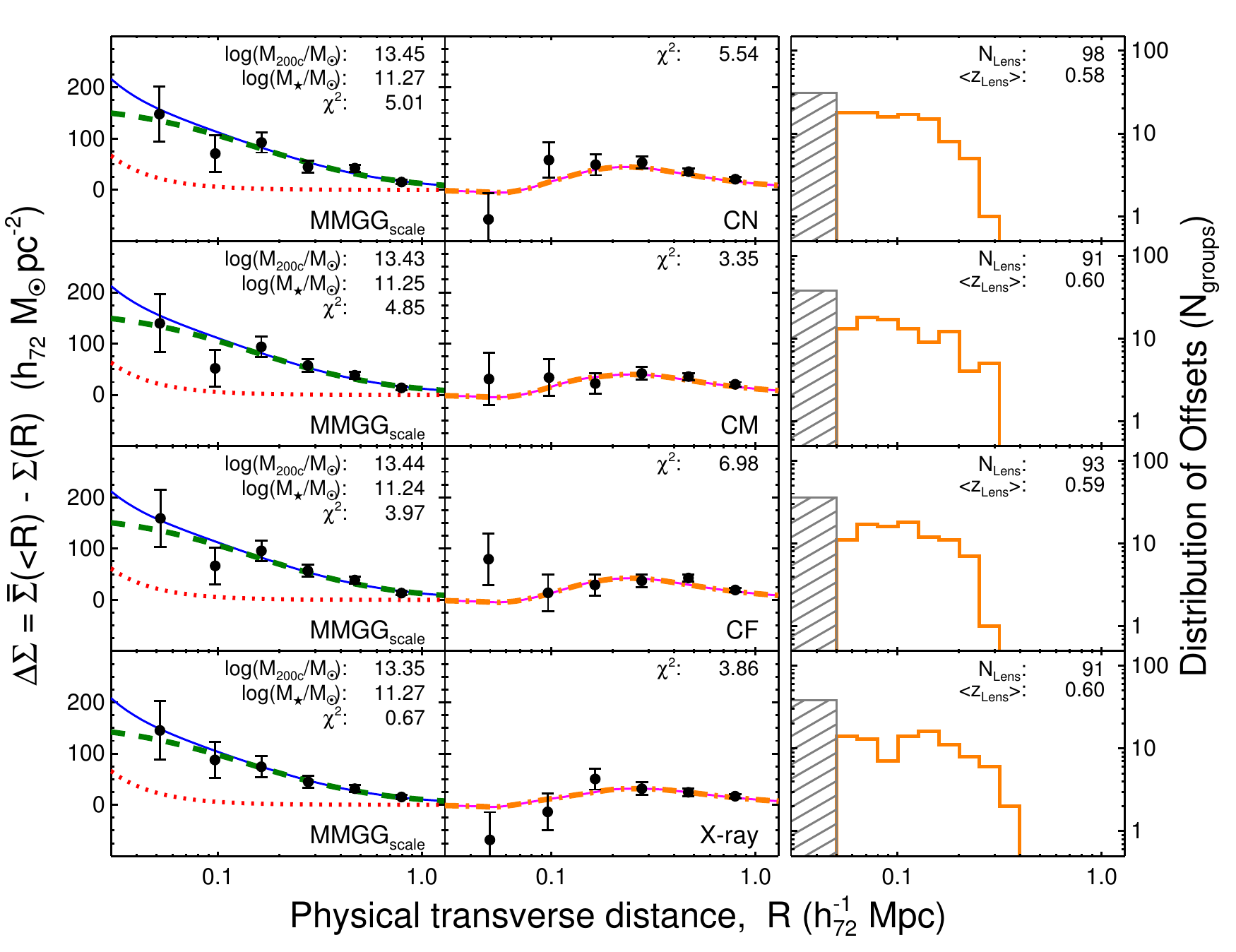}
\caption{Lensing signal for subsets of groups where \textit{centroid
    candidates} are offset from MMGG$_{\rm scale}$ by $>50$~{\rm
    kpc}. The left column shows the signal stacked around MMGG$_{\rm
    scale}$, the middle column shows the signal stacked around an
  alternate candidate, and the right column shows the projected
  distribution of offsets between the two candidate positions. Models are
  discussed in the text with line styles and colors as in Figure~\ref{fig:full_stacks}
  and fit parameters stated within the left column. The
  number of groups with large offsets used in each row is stated in
  the right column, and the gray hashed boxes indicate the number of
  groups excluded from this analysis because the two candidates agree
  to within $50$~{\rm kpc}.}
\label{fig:diff_stacks_centroids}
\end{figure*}

\begin{figure*}[htb]
\plotone{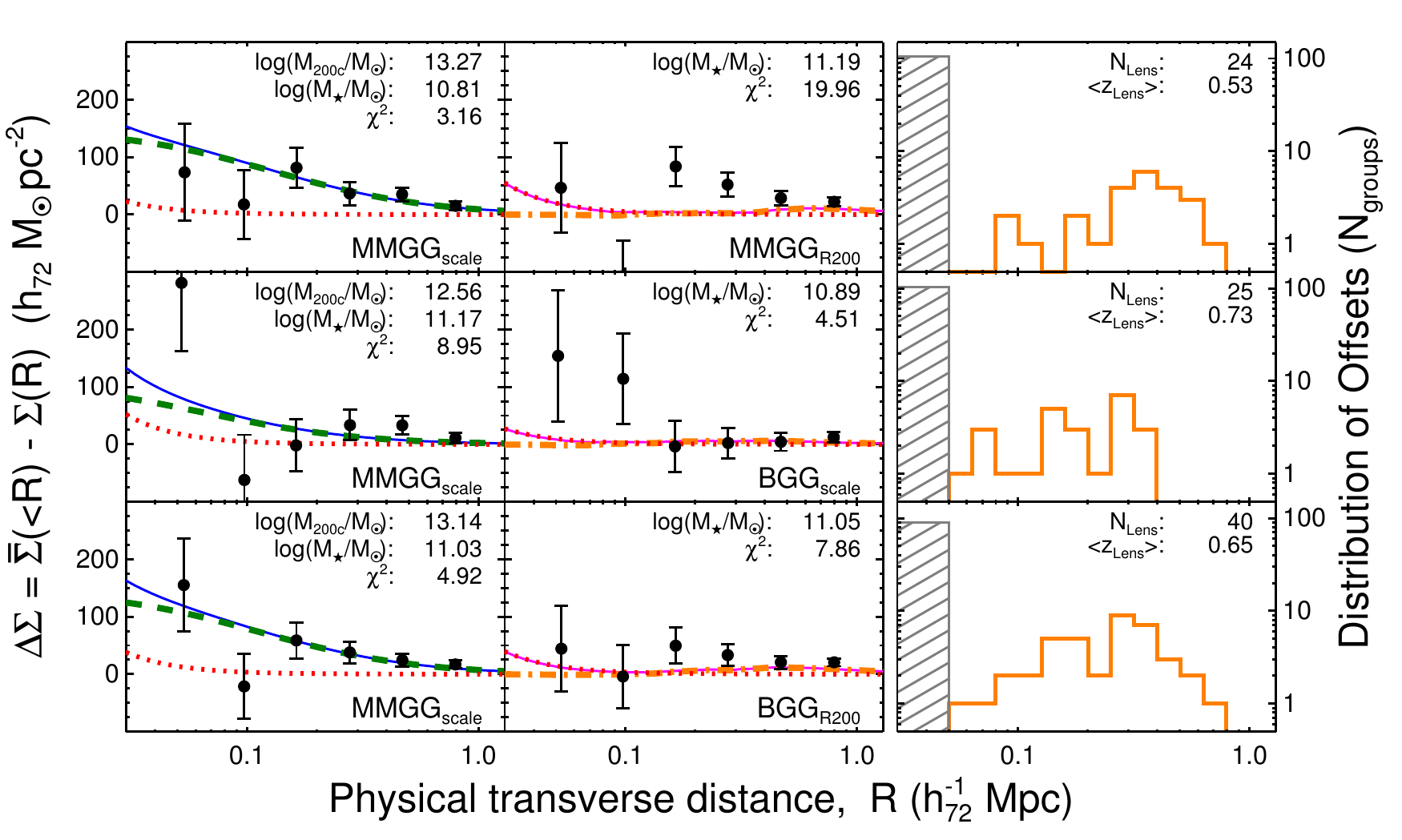}
\caption{Lensing signal for subsets of groups where \textit{galaxy candidates}
  are offset from MMGG$_{\rm scale}$ by $>50$~{\rm kpc}. Plot columns
  and style are the same as in Figure~\ref{fig:diff_stacks_centroids}.}
\label{fig:diff_stacks_galaxies}
\end{figure*}

 While the stacked lensing signal for the full sample of groups shown
in Figure~\ref{fig:full_stacks} exposes differences between candidate
centers, Figure~\ref{fig:offsets} shows that the candidate centers are
identical in many groups. In order to more directly compare different
centering schemes, we now select only the groups in which candidate
centers disagree by a measurable amount. Figures~\ref{fig:diff_stacks_centroids}
and~\ref{fig:diff_stacks_galaxies} present this analysis, with 
each row showing the lensing signal centered on the MMGG$_{\rm scale}$
(left column) and a different candidate (middle column) for the subset
of groups where the two candidate centers are separated by more than
$50$~{\rm kpc}. The rightmost column shows the histogram of offsets
between the two centers repeating Figure~\ref{fig:offsets} on
logarithmic axes. The number of groups and the mean redshift
of each subset are shown at the top right. We have
  chosen the MMGG$_{\rm scale}$ as a fiducial center here since
  Table~\ref{tab:fitPars} suggests that it is among the best
  candidates, but using the BGG$_{\rm scale}$ gives a consistent
  picture.

In each row of the figure, we first model the lensing signal in the
lefthand panel assuming that the MMGG$_{\rm scale}$ correctly traces
the halo center, fitting for the halo mass with concentration
fixed. Next, we take that best-fit centered model and convolve it with
the distribution of offsets in the righthand panel to model the signal
in the middle panel. This is equivalent to replacing the offset
distribution of Equation~\eqref{eq:rayleigh} with the empirical
distribution of offsets between two candidates, leaving no free
parameters for the models shown in the middle panels. We again add a
central point mass fixed to the mean photometric stellar mass of the
candidate galaxies for each of the galaxy
centers. Though we expect that galaxies offset from
  the halo center have subhalos, the lensing data are currently insufficient
  to constrain such an additional contribution. The halo mass is
fitted using only the lensing data centered on the MMGG$_{\rm scale}$,
so the level of agreement between the offset model and the data in the
middle panel can be read as a consistency test.

Looking first at the comparisons between the MMGG$_{\rm scale}$ and
the centroid candidates in Figure~\ref{fig:diff_stacks_centroids}, we see that the lensing signal is
reasonably well-described with an NFW halo centered on the MMGG$_{\rm
  scale}$, with a mean halo mass consistent with that measured for
the full sample in Table~\ref{tab:fitPars}. Additionally, the offset
model gives a decent representation of the lensing signal measured around
the centroid candidates shown in the middle panels. We interpret these
results as indicating that centroid candidates do not trace halo
centers as well as galaxy candidates.

In the samples where the MMGG$_{\rm scale}$ disagrees with other
galaxy candidates (Figure~\ref{fig:diff_stacks_galaxies}), the lensing
signal is different than in the centroid
comparison. The best-fit halo masses are significantly
  lower than in the full sample measured earlier. The lensing signals
  are noisier, in part because of the smaller sample sizes, and one
  case deviates significantly from the offset model.  Though it appears
that the BGG$_{\rm scale}$ produces a higher lensing signal than the
MMGG$_{\rm scale}$ at small radii in this direct comparison, that
profile is not well-fit by a centered NFW model and the fitted mass is
similarly low.

\begin{figure}[htb]
\epsscale{1.3}
\plotone{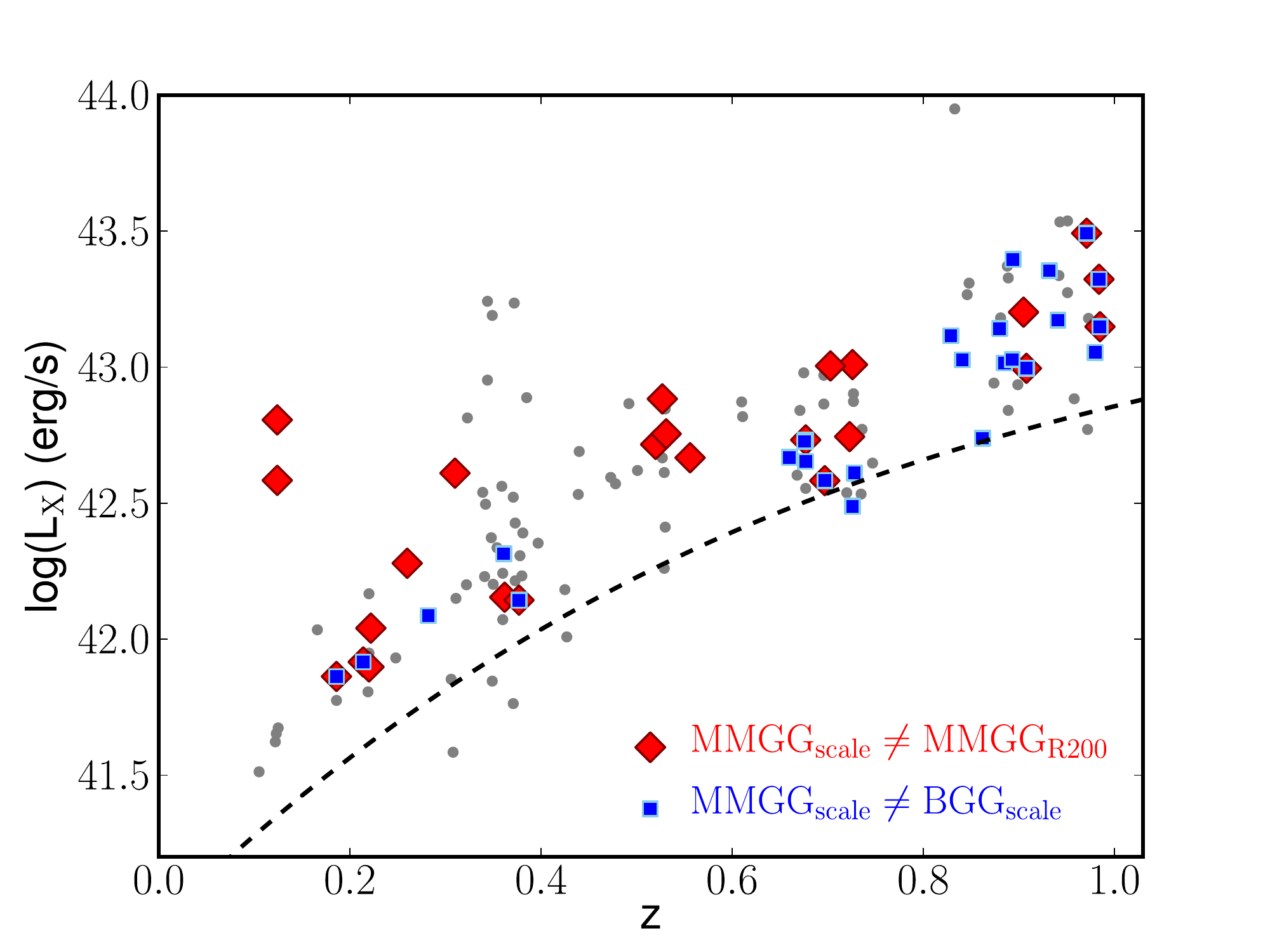}
\caption{Distribution of redshifts and X-ray luminosities for the
  group sample. Cases where different galaxy candidates disagree are
  identified according to the legend. The dashed curve shows the
  $4\sigma$ X-ray flux limit of $1.0\times10^{-15}~\rm{erg cm^{-2}
    s^{-1}}$ reached in $96\%$ of the field converted to a limiting
  luminosity.}
\label{fig:bad_groups}
\end{figure}

Figure~\ref{fig:bad_groups} shows how two samples of groups with
differing galaxy candidates are distributed in redshift and X-ray
luminosity relative to the full sample of groups. Cases where the
most massive group member lies in the outskirts (MMGG$_{\rm scale}
\ne$ MMGG$_{\rm R200}$) appear to be evenly distributed throughout the
sample. Groups where the brightest galaxy near the X-ray position is
not the most massive (MMGG$_{\rm scale} \ne$ BGG$_{\rm scale}$) tend
to be at higher redshifts, which can also be seen in the mean
redshifts for the samples in
Figure~\ref{fig:diff_stacks_galaxies}. This illustrates how an
observed-frame 
selection of BCGs, like that used for BGG$_{\rm scale}$ here, tends to
pick up bluer star-forming galaxies at higher redshifts.

The low masses shown in
Figure~\ref{fig:diff_stacks_galaxies} could be attributed to
statistical fluctuations with the small sample size ($24-40$ groups,
as compared with $91-98$ groups in
Figure~\ref{fig:diff_stacks_centroids}). To test this idea, we performed
jackknife tests using randomly selected samples of the same number of groups without
replacement. In each of $1000$ random samples we measured the lensing
signal around MMGG$_{\rm scale}$ and fit a model with a central point
mass equal to the average stellar mass and a centered NFW with halo
mass as a free parameter and concentration fixed. Of the 1000 random samples of
$24$ groups, $19\%$ had a best-fit halo mass lower than that found for
the sample of $24$ groups where MMGG$_{\rm scale} \ne$ MMGG$_{\rm
  R200}$. In samples of $25$ groups, none had a lower best-fit halo
mass than $\log(M_{\rm 200c}/M_{\odot})=12.57$, the value found for
the sample with MMGG$_{\rm scale} \ne$ BGG$_{\rm scale}$. And in
samples of $40$ groups, only $1\%$ had a best-fit mass lower than the
sample where MMGG$_{\rm scale} \ne$ BGG$_{\rm R200}$. For a given
X-ray luminosity, groups with discrepant galaxy candidate centers
appear to have lower masses than the rest of the sample and
in some cases more
disturbed mass profiles. 

Figure~\ref{fig:bad_groups} shows a redshift dependence in the
fraction of groups where the brightest galaxy (measured from the
observer-frame F814W flux) does not have the highest stellar
mass. High redshift groups may have more disturbed lensing profiles
for different reasons than having ambiguous galaxy center candidates,
so we have repeated the jackknife tests restricting to groups at
$z>0.7$. Even among high redshift systems, the groups with a brighter
galaxy than MMGG$_{\rm scale}$ are outliers; less than $1\%$ of random
samples of groups produce a best-fit mass lower than the samples where
MMGG$_{\rm scale} \ne$ BGG$_{\rm scale}$ or MMGG$_{\rm scale} \ne$
BGG$_{\rm R200}$, while just $17\%$ of random samples are fit by lower
masses than the groups with MMGG$_{\rm scale} \ne$ MMGG$_{\rm R200}$.

Visual inspections of the groups with discrepant galaxy candidates
have not revealed obvious differences from the rest of the
sample. Given the noisy lensing signal measured around such groups, one might
worry that they are not real associations but chance
projections. However, these groups are not preferentially found near
the X-ray flux limit shown in Figure~\ref{fig:bad_groups} and they do
not have unusually low numbers of member galaxies associated with
them, so they are unlikely to be impurities in the group sample.

Groups with multiple massive galaxies and disturbed
density profiles are likely to have merged recently, a point we
discuss further in Section~\ref{s:discussion}.

\subsection{Model Uncertainties}
\label{s:model_unc}

Having established that the lensing data can be used to
constrain the choice of a tracer for the centers of mass of halos, we would like to
know how well a given candidate actually traces halo centers. In this
section we explore a more general model to see how certain
assumptions, namely the mass-concentration relation and the form of
the central mass component, affect our results.

The results shown in the previous section suggest that galaxy
candidates trace the halo centers better than centroid candidates,
because the stacked lensing signal on small scales is greater when
centered on a galaxy. While the model used to fit the signal for
galaxy candidates includes a component to account for the stellar mass
that exists in galaxy candidates but not in centroid candidates, it is
plausible that we have underestimated the amount of mass in the
central galaxies, either in stars and baryons or in a dark matter
subhalo. In a typical halo model the central galaxy has no
subhalo, so in the latter scenario the galaxy candidate could actually be a
satellite or the system may be unrelaxed.
Detailed modeling of strong and weak lensing observations of
individual clusters has found evidence for subhalos around member
galaxies even near cluster centers \citep{Natarajan2007,
  Natarajan2009}. Additional mass in the central term $\Sigma_{\rm
  gal}(R)$ could hide 
significant offsets between the galaxy candidates and host halos,
effectively filling in the decrement at small radii in the halo term
$\Sigma_{\rm halo}(R)$.

To address this degeneracy, we apply our full model with an additional mass
component around the galaxy candidate, in addition to the point source
representing the stellar mass. We use the functional form of a
truncated non-singular isothermal sphere \citep{PastorMira2011}
\begin{equation}
\rho(r)=\frac{\rho_0}{(1+r^2/r_{\rm core}^2)(1+r^2/r_{\rm cut}^2)}.
\label{eq:tnsi}
\end{equation}
We fix the core radius $r_{\rm core}=0.1$~{\rm kpc} and set the truncation
radius $r_{\rm cut}$ to the distance at which the local mass
densities of the subhalo and halo components are equal, as measured
along the line connecting their centers assuming an offset distance
$\sigma_{\rm off}$. To facilitate comparisons
with the stellar mass, we cast the normalization of the model in terms
of a free parameter $M_{\rm sub}$ (instead of $\rho_{0}$) which we
will refer to as the subhalo mass, defined as the mass enclosed within
$5$~{\rm kpc}, the mean half-light radius for our sample. Since the core
radius is much smaller than our innermost lensing measurement, the
exact value chosen does not influence our results. The choice of
truncation radius and functional form of the subhalo mass profile does
affect the mass normalization, but the qualitative shapes of the
parameter degeneracies are consistent for a variety of truncation
schemes and for mass profiles that are isothermal, or nearly so,
across the inner few tens of~{\rm kpc}. However if the subhalo
component is not truncated at all, or if it is modeled instead as an
NFW profile as \citeauthor{PastorMira2011} find in simulations with
subhalos, its mass can become degenerate with the full halo
mass.

For a given lensing signal, a larger halo concentration can also compensate
for some miscentering effects, so we free the $c_{\rm 200c}$ parameter
to study the degeneracy with $\sigma_{\rm off}$ without the
restrictions of the model for concentration used before. In practice,
concentration is not well-constrained by this data set, so a prior is still
needed to restrict the range of values when fitting.

An additional model uncertainty comes from the form of the
distribution of offsets $P(R_{\rm off})$ that is used. The offsets
between galaxy candidates in Figure~\ref{fig:offsets} appear to
support the models used by \citet{Johnston2007b} and \citet{Oguri2010},
with a significant fraction of agreement between candidates along with
a wide tail in the distribution. To avoid the need for additional free
parameters describing both well-centered and miscentered populations in the offset distribution,
we attempt to select a clean sample of groups and constrain the scale
of offsets between the candidate galaxy center and the halo center
assuming the single distribution given in
Equation~\eqref{eq:rayleigh}. We test this model on the sample of
groups for which the four galaxy candidates agree, i.e. where the
brightest and most massive group members are one and the same, and
where this galaxy lies within an NFW scale radius of the X-ray
centroid. There are $85$ groups with such unambiguous galaxy centers.

\begin{figure}[htb]
\epsscale{1.2}
\plotone{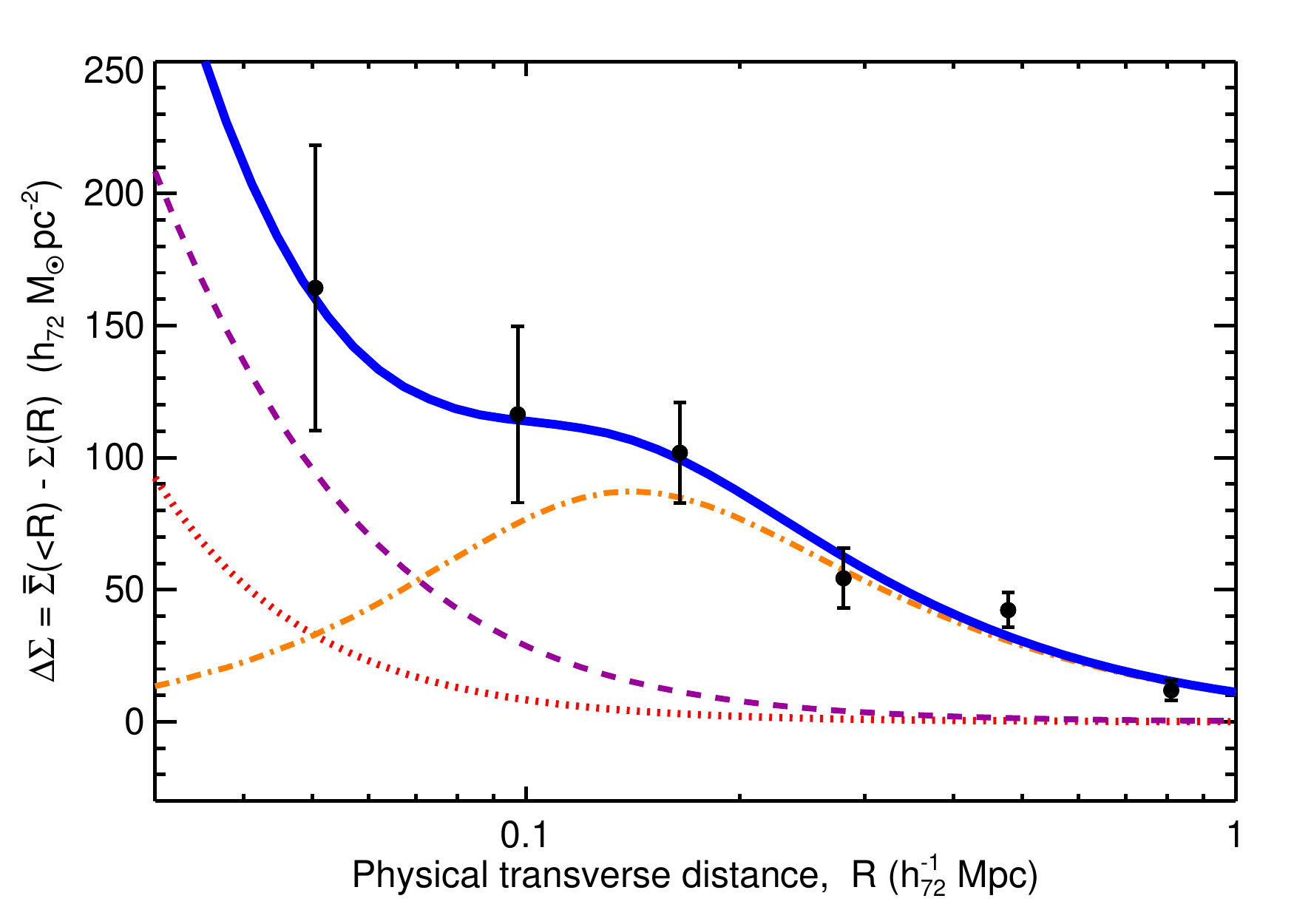}
\caption{Stacked lensing signal for $85$ groups
  where galaxy candidates agree. Model curves show the maximum
  likelihood solution for an offset NFW halo (orange dot-dashed),
  subhalo (purple dashed), and a point source for the stellar
  mass (red dotted), along with the sum (thick solid blue). There are
  degeneracies among some fit parameters; for example, a model with
  smaller centering offsets and a less massive subhalo also fits the data.}
\label{fig:good_centers_ds}
\end{figure}

The lensing signal for these groups stacked around the galaxy
candidate is shown in Figure~\ref{fig:good_centers_ds}. To summarize
our model, we fit the signal with an NFW halo and a truncated
isothermal subhalo of the form in Equation~\eqref{eq:tnsi} in addition
to a point mass fixed by the photometric stellar mass. The point mass is
fixed at the center of the subhalo, which is allowed to be offset
from the center of the halo. This leaves four free parameters:
the halo mass $M_{\rm 200c}$ and concentration $c_{\rm 200c}$, the
offset scale $\sigma_{\rm off}$ between the galaxy and halo, and the
subhalo mass normalization $M_{\rm sub}$.

\begin{figure*}[htb]
\epsscale{0.9}
\plotone{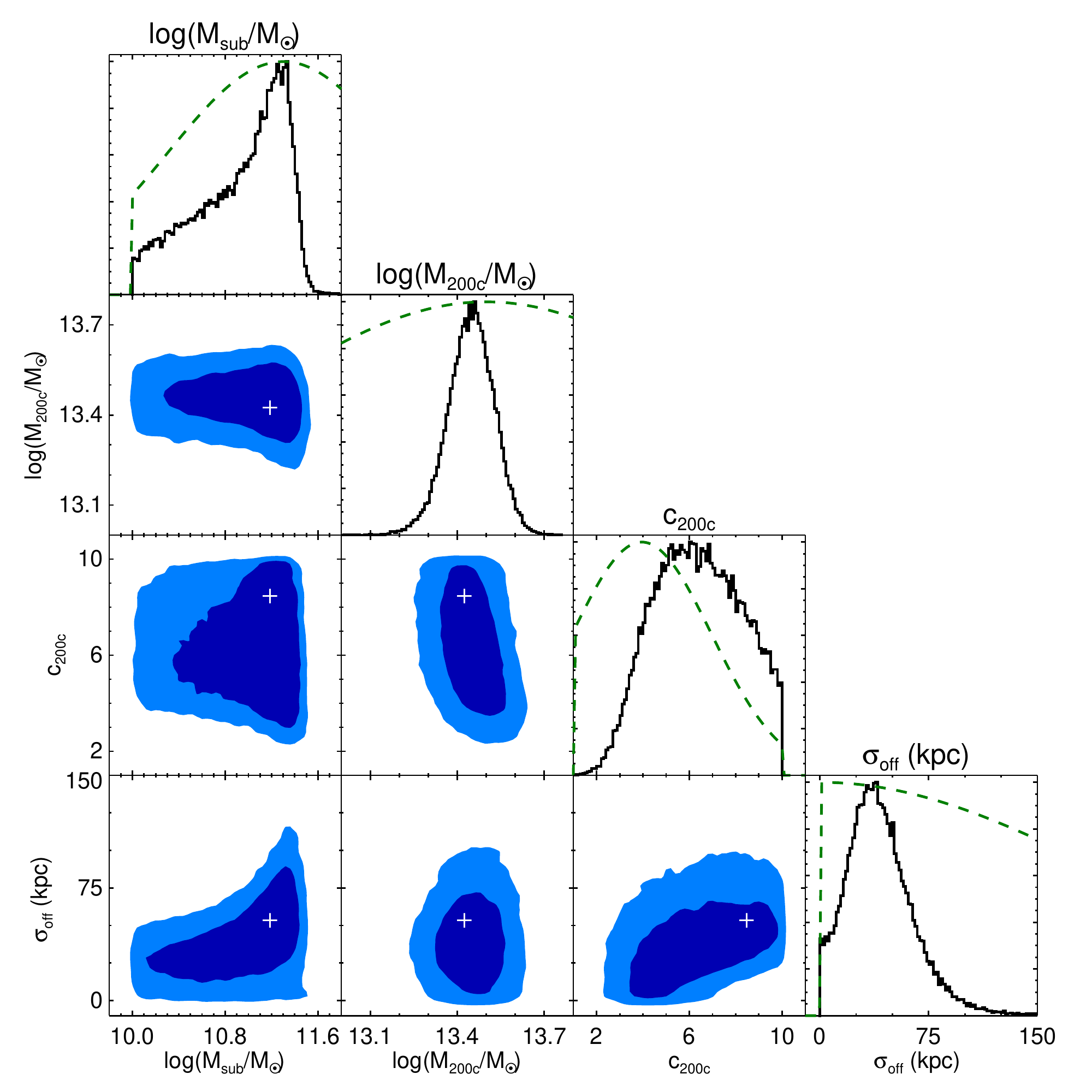}
\caption{Posteriors for the four parameters of a general offset model
  discussed in the text, applied to the lensing signal from
  Figure~\ref{fig:good_centers_ds}. Blue contours show $68\%$ and
  $95\%$ regions for the pair of parameters noted along the axes,
  marginalized over the other parameters. White crosses mark the
  maximum likelihood parameters. Top panels show the
  posterior distributions for single parameters while marginalizing
  over the others (black histograms), along with arbitrarily
  normalized priors (green dashed curves).}
\label{fig:good_centers_cov}
\end{figure*}

Figure~\ref{fig:good_centers_cov} shows the results of the MCMC analysis 
exploring this parameter
space. Each panel with blue contours shows the $68\%$ and $95\%$
regions of the joint posterior probabilities for a pair of parameters,
marginalizing over the other parameters. The top panel in each column
shows for each parameter the arbitrarily normalized prior from
Table~\ref{tab:modelPars} (dashed
green curves) and the one-dimensional posterior probabilities (black
histograms) while marginalizing over the other parameters. Note that
the data are unable to constrain a lower limit on the subhalo mass or
an upper limit to the concentration, so the posterior distributions
are sensitive to our priors in those regions.

The data are
consistent with a model in which there are no subhalos around the
candidate central galaxies, though a value comparable to the average
stellar mass of the galaxies is preferred, and significantly more
massive subhalos are ruled out. We emphasize the stellar mass is
already included in the model as a point source, so the subhalo
component represents mass in excess of the stellar mass that was estimated from
the photometry. Recall that the $M_{\rm sub}$ is
defined as the subhalo mass within the typical half-light radius of
$5$~{\rm kpc}; integration of Equation~\eqref{eq:tnsi} out to large
radii for the maximum likelihood parameters gives a \textit{total}
subhalo mass of roughly five times the observed stellar mass component. The
truncation radius for this model is $21$~{\rm kpc}.  The preference for a nonzero subhalo
mass may simply indicate that the photometric stellar mass is
underestimated\footnote{Stellar masses used in this work have been
  derived assuming a \citet{Chabrier2003} initial mass function
  (IMF). Recent suggestions of a steeper subsolar IMF for early type galaxies
  could explain this discrepancy
  \citep[e.g.,][]{Auger2010,vanDokkum2010}.}, or that the point source
model for the stellar term is insufficient. A subset of
  correctly-centered groups (or an excess at small scales in the offset
  distribution relative to the model) could also explain the signal which we have fitted as a
  subhalo component. The halo mass is
well-constrained and consistent with the values measured for the full
group sample with the simpler model in Section~\ref{s:comparing_cen}.
The data prefer a higher
value of $c_{\rm 200c}$ than the \citet{Zhao2009} prediction for this
halo mass and redshift, but the concentration is not
well-constrained. For the offset distribution, the data are consistent
with no offsets between the central galaxy candidates and the halo
centers, though a value of roughly $50$~{\rm kpc} is preferred. Large
offsets ($\sigma\gtrsim100$~{\rm kpc}) are ruled out for an offset
model of a single two-dimensional gaussian distribution, though
individual outliers may exist. On average, massive galaxies trace halo
centers quite well.

There are degeneracies between the offset distance scale and the
subhalo mass as well as the halo concentration. If the subhalo mass
within $5$~{\rm kpc} is smaller than $\sim10^{11} M_{\odot}$, the
offset scale must be smaller than $50$~{\rm kpc}, but for larger
subhalo masses the offset scale can reach $100$~{\rm kpc}. A
constraint on the total mass enclosed within small scales, perhaps
from strong lensing or measurements of stellar velocity dispersions
for the central galaxies, could improve constraints on $\sigma_{\rm
  off}$. Similarly, $c_{\rm 200c}$ and $\sigma_{\rm off}$ are
correlated, with higher concentrations corresponding to larger
offsets. Thus when modeling a given lensing signal, if miscentering is
not taken into account it can lead to an underestimate of the
concentration.


\section{Discussion}
\label{s:discussion}

Finding the centers of mass for dark matter halos presents both an
observational challenge and an 
opportunity to study the interplay between halos and central
galaxies. In this paper, we have presented a method to test different
tracers of group and cluster centers by comparing the weak lensing
signal stacked around their positions. Our approach can in principle
be applied to any analysis that would be affected by miscentering. For
instance, centering algorithms can also be tested using satellite
kinematics by
identifying the candidate that is nearest to the dynamical center when
averaging over the ensemble. The spatial clustering of galaxies can
also provide centering information; in practice the peaks in the
galaxy density field are generally used in defining optical cluster
catalogs, but clustering data could also be used to optimize the
determination of their centers. Additionally, these approaches to
optimizing the centers of a halo catalog can be fed back into
algorithms used to find halos and their centers, allowing a more
probabilistic approach to deal with cases where centering is uncertain.

With X-ray detected galaxy groups in the COSMOS field, we find that individual
bright and massive galaxies trace the centers of halos better than the
nominal X-ray centroid or the mean position of group members, even
when weighted by luminosity or stellar mass. Offsets between the X-ray
centroids and candidate central galaxies are primarily due to
the large uncertainties in the X-ray positions,
which can reach roughly $200$~{\rm kpc}. Centroids based on the mean
positions of galaxies also have significant uncertainties, though some
offsets exceed the estimated errors. More stringent constraints on
intrinsic offsets between centroid candidates and halo centers could
be obtained with deeper, high-resolution X-ray or SZ data, or with a
sample of massive clusters and more member galaxies.

For each pair of galaxy candidates defined in Section~\ref{s:centers},
roughly $20-30\%$ of groups have ambiguous centers. Either the most
massive galaxy lies far from the X-ray position, or the brightest
galaxy is not the most massive. Only two-thirds ($85$ out of $129$) of the
groups show complete agreement between all four galaxy candidates,
which underlines the importance of testing any choice of center in a group
or cluster catalog.

The results of Figure~\ref{fig:diff_stacks_galaxies} and the jackknife
tests described in Section~\ref{s:discrepant} suggest that groups with
ambiguous galaxy centers have low masses for their X-ray
  luminosities or disturbed halo mass profiles. A
connection between centering offsets and the dynamical state of halos
has been seen in observational studies \citep[e.g.,][]{Forman1982,
  Katayama2003, Sanderson2009} and simulations
\citep[e.g.,][]{Cohn2005, Poole2006, Maccio2007, Neto2007,
  SkibbaMaccio2011}. The ability to identify unusual
  or unrelaxed groups and
clusters with a simple observable such as having different candidate
centers could prove useful when trying to select a clean sample of
relaxed systems. Further studies connecting lensing measurements with
member galaxy properties such as the distribution of colors may
provide a clearer picture of the impact of halo mergers on galaxies
and star formation. Similarly, a more general analysis of the gaps in
stellar mass or luminosity and the spatial separation between massive
member galaxies would complement the restricted set of center
candidates studied here.

There are several reasons why candidate centers can be offset from
halo centers. Interlopers or incompleteness in the group member sample
can result in the wrong galaxies being selected. Satellite galaxies
can be more massive or luminous than centrals due to scatter in the
relation linking the observable property to the mass of a halo or
subhalo. In Paper I, we presented tests of our group membership and
centering algorithm on mock catalogs designed to reflect the real
uncertainties in X-ray positions and photometric redshifts. In that
analysis, $77\%$ of central galaxies were correctly identified as the
MMGG$_{\rm scale}$.  The most common failure mode was the selection of
satellites as MMGG$_{\rm scale}$ because they had higher stellar
masses than the true centrals, which happened in $12\%$ of
cases. The frequency of this occurrence depends on the
  way we populate halos with mock galaxies following the stellar
  mass-halo mass relation of \citet{Leauthaud2012}, but we note that
  scatter in this relation is consistent with that found in other
  analyses \citep{Yang2009, More2009, Reddick2012}. The remaining
centering failures were evenly split between cases where X-ray errors
misplaced the search region and photometric redshift errors scattered
centrals out of their groups.

Observational uncertainties, such as scatter in stellar mass estimates
($\sim0.2$~{\rm dex}) and catastrophic errors in photometric redshifts
(for $\lesssim$ a few percent of objects in our COSMOS sample), add to
this scatter and increase the chances that a satellite is incorrectly
identified as the central galaxy. Furthermore, merging activity is not
modeled in the mocks catalogs, so these effects together can account
for the $\sim30\%$ of observed groups where galaxy candidates
disagree. The offsets seen between centroid candidates (CN, CM, CF,
and X-ray) are consistent with being due primarily to the large
observational uncertainties in their positions.

The offsets measured in this paper are significantly smaller than the
distribution measured from mock catalogs by
\citet{Johnston2007b}. This could be due in part to our choice of
offset model; in Sections~\ref{s:comparing_cen} and \ref{s:model_unc}
we assumed a single Rayleigh distribution of offsets, whereas
\citeauthor{Johnston2007b} separated groups that were correctly
centered from those that were miscentered. Our aim in
Section~\ref{s:comparing_cen} was simply to estimate a typical offset
scale for comparing candidate centers, and in
Section~\ref{s:model_unc} we selected a sample where centering seemed
unambiguous to study up close the offsets of massive galaxies which
could be sloshing around halo centers. The offsets found in
Section~\ref{s:model_unc} are comparable to the smaller component of
the offset distribution measured by \citet{Oguri2010}. While they
are fairly small, modeling degeneracies with the subhalo mass and halo
concentration increase their uncertainty. Future analyses of halo
properties may benefit from including a range of models for the
distribution of offsets, if the data can constrain a larger number of
parameters.

Another important difference between our analysis and that of
\citeauthor{Johnston2007b} is that the maxBCG clusters are detected as
optical galaxy overdensities, whereas the COSMOS groups studied here
are detected in X-ray emission, which traces the dense regions near
the centers of halos, offering a better starting point for finding the
centers of halos. When comparing the effects of miscentering for
different cluster catalogs, it is also worth noting that more massive
clusters are larger, so offsets of a given distance can be more easily
detected than in less massive systems.

Further investigations with simulations could improve our understanding of some issues
with centering \citep[e.g.,][]{Maccio2007, Neto2007, Hilbert2010, Behroozi2011,
  Dietrich2012, Power2012}. There is a similar ambiguity when defining
the center of a simulated dark matter halo, since there can be offsets
between the position of the most bound particle, the mass density
peak, and the centroid of mass with a given smoothing scale. Offsets
between these positions in simulations have been shown to correlate
with the dynamical state of a halo, with larger offsets seen in less
relaxed halos that have experienced a recent merger. Projecting the
matter density in simulated halos to compare with the observed lensing
signals could help explain the poor fits to NFW profiles and unusually
low masses we obtain from samples with ambiguous centers. Similarly, a
better understanding of the form and evolution of halo and subhalo mass
profiles, due to effects like gas cooling or heating and tidal
stripping, could improve our modeling.

Finally, we consider the implications of miscentering for cosmological
analyses. The abundance of massive halos is sensitive to the amplitude
of matter fluctuations and the growth history of the Universe, and the
precise determination of group and cluster masses is a critical aspect
of this probe. \citet{Mandelbaum2010} have studied the effects of
miscentering on mass estimates for massive halos using simulated data
and analytical profiles, assuming the distribution of offsets from the
mock catalogs of \citet{Johnston2007b}. For clusters at the upper end
of our mass range ($\sim10^{14}~\rm{M}_{\odot}$), they find that the
weak lensing mass is underestimated by $25-30\%$ if miscentering is
ignored, and that this effect is stronger for less massive halos (see
their Figure 3). They also show that the accuracy of mass estimates
depends on the assumed concentration as well as the inner and outer
radius measured, and suggest excising the inner regions from the
analysis because of these uncertainties. In a separate study,
\citet{Mandelbaum2008} measured the mass-concentration relation for
several samples of galaxies and clusters, including the maxBCG
sample. Miscentering of clusters is also a concern for measuring
concentrations, producing values that are biased low if the effect is
ignored. \citet{Mandelbaum2008} argued that the miscentering
distribution from \citet{Johnston2007b} may be overestimated, based on
the concentrations they derived from lensing and by comparison with the
distribution of offsets between BCGs and X-ray peaks in a subsample of
clusters \citep{Koester2007b}. Still, they had to trade off statistical
power by excluding data at small radii to reduce systematic
uncertainties from modeling the distribution of offsets.

In this paper, we have endeavored to improve the accuracy of finding
halo centers and to provide constraints on the distribution of offsets
between observational tracers and underlying mass centers. With more
accurate centers (and uncertainties on those positions), we can more
reliably use data at small radii and improve statistical constraints
from group and cluster surveys. Table~\ref{tab:fitPars} suggests
that the halo mass inferred from the lensing signal stacked around our
galaxy candidates is affected at the $5-10\%$ level if miscentering is
ignored. This is significantly smaller than the bias seen by
\citet{Mandelbaum2010} despite the trend that masses are increasingly
underestimated for lower mass halos. We attribute the difference to
the smaller offsets seen in this sample compared to the miscentering
distribution used by \citet{Johnston2007b} where a fraction of groups
had a distribution of offsets with $\sigma_{\rm off}=420 h^{-1}~{\rm
  kpc}$. However, we also use a different model for the offset
distribution and can see in Table~\ref{tab:fitPars} that the masses
estimated from centroid candidates are more biased ($15-30\%$) when
miscentering is not addressed in the model. The statistical
uncertainties on our halo mass estimates are still comparable to the
centering bias, and with this sample we are currently unable to put a
significant constraint on the halo concentration. But with larger
group and cluster samples, our approaches to optimizing the choice of
halo centers and modeling the distribution of offsets will enable
better constraints on halo mass profiles for both astrophysical and
cosmological applications.


\section{Summary and Conclusions}
\label{s:conclusion}

We summarize the main results of this paper as follows:
\begin{enumerate}
\item In our data set, different definitions of group centers do not
  always agree and occasionally show large offsets from one another. Candidate centers
  based on the locations of the brightest or most massive galaxy
  differ in 20-30\% of cases with a wide range of offsets. Centroids based on the
  mean position of member galaxies or X-ray flux have offsets
  from other center definitions that are roughly
  consistent with their larger statistical uncertainties ($\sim50-150~{\rm kpc}$).
\item The offsets between centers produce a measurable signal in the
  lensing profile. Stacking the signal around a bright or massive
  galaxy tends to produce a larger signal at small scales than any
  centroid not located on a galaxy, and the difference in
  these signals is greater than expected from the stellar mass of the galaxy.
\item Among the candidate centers we have tested, the brightest or the
  most massive galaxy near the X-ray centroid appear to be the best
  tracers of the center of mass of halos. Centering definitions based
  on the centroid of member galaxies have larger offsets and uncertainties.
\item Groups that have ambiguous centers because of multiple bright or
  massive galaxies have lensing signals that suggest a lower mass than expected given their X-ray
  luminosity and in some cases appear
    disturbed. These are possibly merging systems, and the property
  of having discrepant candidate centers gives a simple observational
  indicator to identify them.
\item In groups with a clear central galaxy, offsets between the
  galaxy and the halo center are fairly small ($\lesssim 75~{\rm
    kpc}$).  The offset is somewhat degenerate with the amount of
  substructure around the galaxy and with the concentration of the
  group halo.
\end{enumerate}

These findings apply to our group sample but our approach can readily
be applied to other group and cluster data sets, and to different
analyses such as satellite dynamics and richness-based mass
estimators. Given the level of disagreement among our candidate
centers, we advise testing different centers to determine the degree
to which centering choices affect a given analysis. Our finding that groups
with ambiguous centers are less massive for their
  X-ray luminosity or have disturbed mass profiles suggests that
these groups could be identified and excluded if an analysis calls for
a clean sample of halos. Additionally, a probabilistic approach to 
centering algorithms could provide information about the confidence in
a given center allowing for appropriate weighting of different systems.

Much larger samples of groups and clusters are being constructed with
ongoing and upcoming surveys such as the South Pole Telescope and
Atacama Cosmology Telescope, the Dark Energy Survey, and eROSITA. This
initial study with a modestly-sized sample of groups benefits from
high-resolution X-ray selection which provides a good starting point for finding
centers, but a similar approach can be applied to optical or SZ-selected
catalogs. With larger samples, we can hope to improve constraints on
the density profiles of dark matter halos including their
concentration and inner slope, as well as the mass distribution in
subhalos and the effects of mergers. Additional constraints on the
mass distribution from strong lensing or stellar kinematics can
provide interesting constraints on a range of scales and improve
models for the distribution of offsets between galaxies and halo
centers.

\acknowledgments 
We thank Uros Seljak and Priya Natarajan for
constructive comments on a draft of the paper, as well as Ami Choi and
David Schlegel for helpful discussions. MRG has been
supported by the US Department of Energy's Office of High Energy
Physics (DE-AC02-05CH11231) and a Graduate Research Fellowship from
the US National Science Foundation. This work was also supported by World Premier
International Research Center Initiative (WPI Initiative), MEXT,
Japan.

We gratefully acknowledge the contributions of the entire COSMOS
collaboration consisting of more than 70 scientists. More information
on the COSMOS survey is available at {\bf
  \url{http://www.astro.caltech.edu/~cosmos}}.  This research has made
use of the NASA/IPAC Infrared Science Archive, which is operated by
the Jet Propulsion Laboratory, California Institute of Technology,
under contract with the National Aeronautics and Space Administration.

\mbox{~} 



\bibliographystyle{apj}

\bibliography{centering}


\end{document}